\documentclass[final,5p,times,twocolumn]{elsarticle}

\usepackage{graphicx}

\usepackage{amssymb}

\usepackage{lineno,hyperref}
\modulolinenumbers[5]

\journal{XYZ}

\begin{document}

\begin{frontmatter}

\title{Revealing neutral bremsstrahlung in two-phase argon electroluminescence}





\author[A,B]{A.~Buzulutskov\corref{corresponding author}}
\cortext[corresponding author]{Corresponding author.}
\ead{A.F.Buzulutskov@inp.nsk.su}
\author[A,B]{E.~Shemyakina} 
\author[A,B]{A.~Bondar}
\author[B]{A.~Dolgov} 
\author[A,B]{E.~Frolov}
\author[A,B]{V.~Nosov}
\author[A,B]{V.~Oleynikov}
\author[A,B]{L.~Shekhtman}
\author[A,B]{A.~Sokolov}
\address[A]{Budker Institute of Nuclear Physics SB RAS, Lavrentiev avenue 11, 630090 Novosibirsk, Russia}
\address[B]{Novosibirsk State University, Pirogov street 2, 630090 Novosibirsk, Russia}

\begin{abstract}
Proportional electroluminescence (EL) in noble gases has long been used in two-phase detectors for dark matter search, to record ionization signals induced by particle scattering in the noble-gas liquid (S2 signals).  Until  recently, it was believed that proportional electroluminescence was fully due to VUV emission of noble gas excimers produced in atomic collisions with excited atoms, the latter being in turn produced by drifting electrons. In this work we consider an additional mechanism of proportional electroluminescence, namely that of bremsstrahlung of drifting electrons scattered on neutral atoms (so-called neutral bremsstrahlung); it is systemically studied here both theoretically and experimentally. In particular, the absolute EL yield has for the first time been measured in pure gaseous argon in the two-phase mode, using a dedicated two-phase detector with EL gap optically read out by cryogenic PMTs and SiPMs.  We show that the neutral bremsstrahlung effect can explain two intriguing observations in EL radiation: that of the substantial contribution of the non-VUV spectral component, extending from the UV to NIR, and that of the photon emission at lower electric fields, below the Ar excitation threshold. Possible applications of neutral bremsstrahlung effect in two-phase dark matter detectors are discussed.
\end{abstract}

\begin{keyword}
dark matter detectors; two-phase detectors; liquid argon; proportional electroluminescence; neutral bremsstrahlung
\end{keyword}

\end{frontmatter}


\section{Introduction}

Proportional electroluminescence (EL) in noble gases (or differently secondary scintillation) is the effect that is routinely used in two-phase (liquid-gas) detectors to record ionization signals in the gas phase, induced by particle scattering in the liquid phase  (so-called S2 signals) \cite{Aprile06,Bolozdynya10,Chepel13, Bernabei15}. Such two-phase detectors are relevant for dark matter search \cite{Xenon12,Lux13,Panda14,DarkSide15,DarkSide20k} and low energy neutrino detection \cite{CoNu04,CoNu09,Red17}.  The S2 signals are recorded typically in the EL gap placed above the liquid-gas interface, optically read out using either cryogenic PMTs \cite{Xenon12,Lux13,Panda14,DarkSide15} or cryogenic SiPMs \cite{DarkSide20k}. 

Until  recently, it was commonly believed that proportional electroluminescence was fully due to vacuum ultraviolet (VUV) emission of noble gas excimers, e.g. Ar$^{\ast}_2(^{1,3}\Sigma^+_u)$, produced in three-body atomic collisions with excited atoms, e.g.  Ar$^{\ast}(3p^54s^1)$, which in turn are produced by drifting electrons in electron-atom collisions: see review \cite{ArXeN2Proc17}.

While proportional electroluminescence in heavy noble gases (Ar, Kr and Xe) was thoroughly studied at room temperature, both theoretically and experimentally \cite{ArELExp08,ArELTheory11}, little was known about that at cryogenic temperatures. The study of proportional electroluminescence in gaseous Ar at cryogenic temperatures, in the two-phase mode, has been recently started by our (Novosibirsk) group \cite{CRADPropEL15,CRADELGap17,CRADPropEL17}. The first results were obtained in argon with a minor (10-50 ppm) admixture of nitrogen. The idea to use the N$_2$ dopant was to shift the VUV emission of Ar to the near ultraviolet (UV) emission of N$_2$ directly in the detection medium, in order to avoid using a wavelength  shifter (WLS) in front of PMTs or SiPMs.  It relied on the fact that in the presence of N$_2$, the excimer production (and hence the VUV emission) can be taken over by that of excited N$_2$ molecules (due to excitation transfer from Ar to N$_2$ species), followed by their de-excitations in the near UV through the emission of the so-called Second Positive System (2PS), at 310-430 nm \cite{ArXeN2Proc17,Takahashi83}. 

However, the results obtained in \cite{CRADPropEL15,CRADPropEL17} have been mixed. On one hand, the presence of the non-VUV component (i.e. that of UV and visible) was confirmed in proportional electroluminescence in two-phase Ar doped with N$_2$. On the other hand, at such small N$_2$ contents (10-50 ppm) the appearance of the non-VUV component could not be explained in a simple model of excitation transfer from Ar to N$_2$ \cite{ArXeN2Proc17}. Hence, the problem of proportional electroluminescence in two-phase Ar remained unresolved. 

In this work, we partially resolved the problem: we have studied proportional electroluminescence in pure gaseous Ar, without additives, in the two-phase mode. Surprisingly, we again observed a non-VUV component in EL radiation. Moreover, this component was observed not only at higher electric fields, but also at lower fields, below the Ar excitation threshold. 
These unexpected observations made us recall the idea of an additional EL mechanism in two-phase Ar, namely that of neutral bremsstrahlung (NBrS), that acts concurrently with the ordinary mechanism.

Bremsstrahlung of drifting electrons scattered on neutral atoms (so-called neutral bremsstrahlung) is produced by slow electrons, at the energies of the order of 1-10 eV \cite{Firsov61,Kasyanov65,Dalgarno66,Biberman67,Kasyanov78}. 
The NBrS effect was used to explain continuous emission spectra in a weakly ionized plasma \cite{Batenin72,Rutscher76,Park00}. The NBrS effect was also used to develop a detection technique for ultra-high-energy cosmic rays, where the NBrS radiation (in the radio-frequency range) is emitted by low-energy ionization electrons left after the passage of the showers in the atmosphere \cite{AlSamarai16}.  

More interesting is that back in 1970, the NBrS mechanism was proposed as an explanation for proportional electroluminescence in xenon \cite{Butikov70}. In subsequent work \cite{DeMunari71}, the photon emission rate was theoretically calculated for NBrS electroluminescence; it was stated there that in Xe the EL rate agrees, in order of magnitude, with the calculated NBrS rate of light production.  

However this statement was refuted in \cite{Dias86}. Here is a quote from that work: "The agreement of our results with previous experimental work shows that mechanisms based on the direct excitation of the noble gas atoms by the electrons can fully account for the secondary light production. There is no need to consider other processes like ... neutral bremsstrahlung". 
Since then bremsstrahlung in connection with electroluminescence was almost forgotten. It was mentioned only once as a hypothesis explaining the weak under-threshold electroluminescence observed in gaseous krypton  \cite{Barabash93}, with reference to the unpublished work \cite{Bolozdynya85}. 

In the present work we show that at least for Ar the NBrS mechanism is needed to explain the properties of proportional electroluminescence: we developed the theory of NBrS electroluminescence that for the first time quantitatively described the experiment at lower fields, below the excitation threshold, and that has chances to do it at higher fields, above the threshold. Accordingly, this work might be considered as a revival of the idea of NBrS electroluminescence.

The present study was performed in the course of the development of two-phase Cryogenic Avalanche Detectors (CRADs) of ultimate sensitivity for rare-event experiments \cite{CRADRev12,CryoMPPC15,XRayYield16,IonYield17}.

\section{Theory of neutral bremsstrahlung in proportional electroluminescence}

\begin{figure}[!htb]
	\centering
	\includegraphics[width=0.99\columnwidth,keepaspectratio]{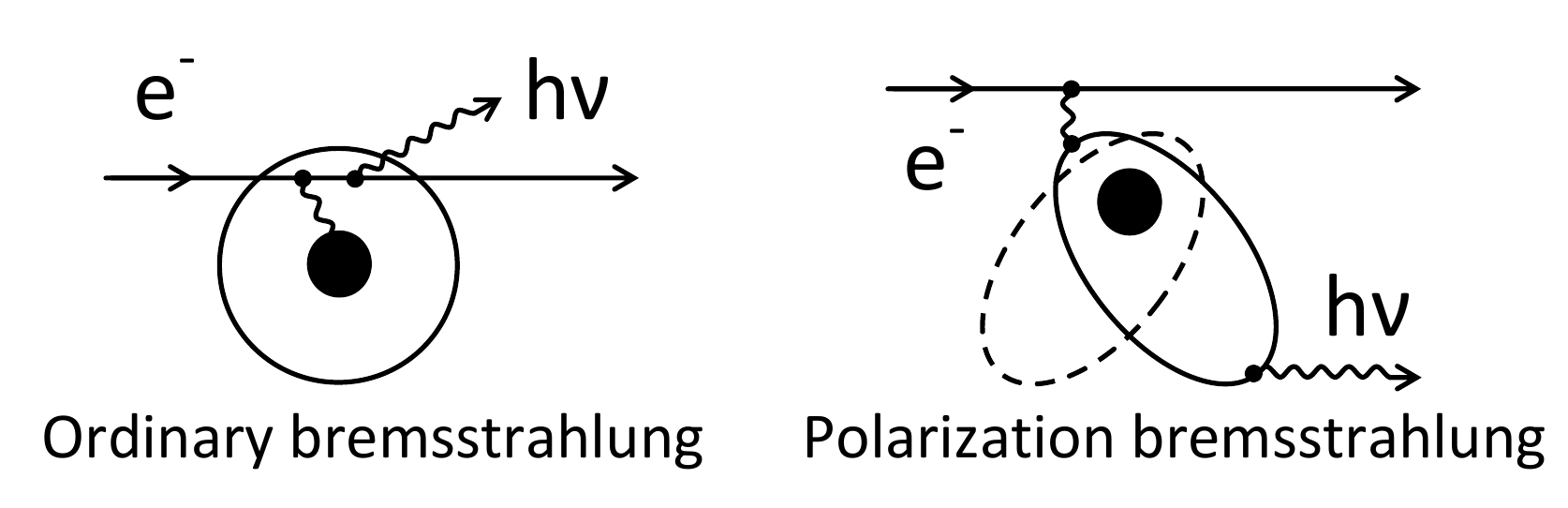}
	\caption{Schematic representation of the ordinary and polarization bremsstrahlung processes.}
	\label{OBrS}
\end{figure}

\begin{figure}[!htb]
	\centering
	\includegraphics[width=0.99\columnwidth,keepaspectratio]{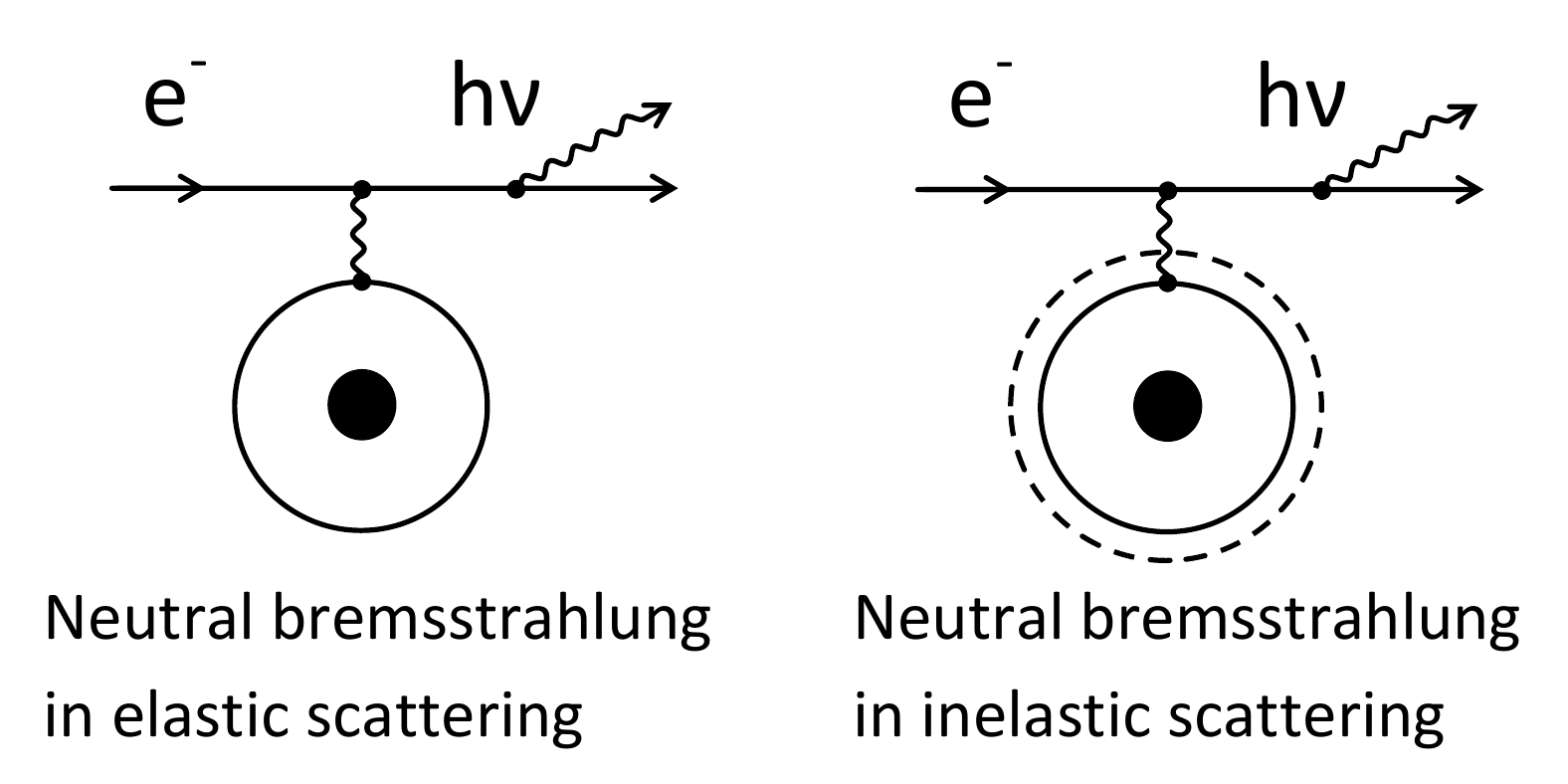}
	\caption{Schematic representation of the neutral bremsstrahlung process in elastic and inelastic electron scattering.}
	\label{NBrS}
\end{figure}

Bremsstrahlung of electrons on nuclei (ordinary bremsstrahlung or OBrS) may be considered as the most famous electromagnetic process involving electrons and photons. There are two other less known electron bremsstrahlung processes: that of polarization bremsstrahlung (PBrS) and that of neutral bremsstrahlung (NBrS). These are schematically depicted in Figs.~\ref{OBrS} and \ref{NBrS}. 

Polarization bremsstrahlung is produced by atoms due to their time-dependent polarization, induced by fast electrons when scattered on atoms: see Fig.~\ref{OBrS} and review \cite{PBrS14}. At electron energies of the order of 1 keV, its intensity is predicted to be comparable with that of ordinary bremsstrahlung.

Neutral bremsstrahlung is produced by slow electrons when they are scattered (elastically or inelastically) on neutral atoms (Fig.~\ref{NBrS}):
\begin{eqnarray}
\label{Rea-NBrS-el}
e^- + A \rightarrow e^- + A + h\nu \; , \\
\label{Rea-NBrS-exc}
e^- + A \rightarrow e^- + A^{\ast} + h\nu \;.
\end{eqnarray}
We will see in the following that NBrS spectrum lies in the UV, visible and near infrared (NIR) regions and that its intensity is maximal for elastic collisions (reaction \ref{Rea-NBrS-el}), at electron energies of the order of 1-10 eV. At such energies the contribution of ordinary and polarization bremsstrahlung can be neglected, since the OBrS emission is suppressed due screening of the Coulomb field of the atomic nucleus, and since the PBrS intensity was shown to be two orders of magnitude smaller than that of NBrS \cite{Kasyanov65}. 

In the general case, the electron-atom scattering producing a NBrS photon may be accompanied by excitation of an atom (reaction \ref{Rea-NBrS-exc}). For this case, the analytical formula for differential cross section of NBrS photon emission was given in \cite{Kasyanov78} (in the approximation of an even partial wave): 
\begin{eqnarray}
\label{Eq-sigma-exc} 
\left( \frac{d\sigma}{d\nu} \right)_{NBrS,exc} = \frac{8}{3} \frac{r_e}{c}  \frac{1}{h\nu} \times \hspace{120pt} \nonumber \\ 
\times \left[\left( \frac{E - E_{exc} - h\nu}{E-E_{exc}} \right)^{1/2}    
(E-E_{exc} - h\nu) \ \sigma _{exc}(E) \ + \nonumber \right. \\
\left. + \ \left( \frac{E - h\nu}{E} \right)^{1/2} E \ \sigma _{exc}(E - h\nu) \right] \; . 
\end{eqnarray}
Here $r_e=e^2/m_e c^2$ is the classical electron radius, $c=\nu \lambda$ is the speed of light, $E$ is the initial electron energy, $E_{exc}$ is the atom excitation energy,  $h\nu$ is the photon energy, and $\sigma _{exc}(E) $ is the electron inelastic (excitation) cross section as a function of electron energy. 

For elastic electron-atom scattering $E_{exc}=0$ and the differential cross section for NBrS photon emission is reduced to a more familiar formula, described in a number of works \cite{Firsov61,Kasyanov65,Dalgarno66,Biberman67,Park00}:
\begin{eqnarray}
\label{Eq-sigma-el} 
\left( \frac{d\sigma}{d\nu} \right)_{NBrS,el} = \frac{8}{3} \frac{r_e}{c} \frac{1}{h\nu} \left(\frac{E - h\nu}{E} \right)^{1/2} \times \hspace{40pt} \nonumber \\ \times \ [(E-h\nu) \ \sigma _{el}(E) \ + \ E \ \sigma _{el}(E - h\nu) ]  \; ,
\end{eqnarray}
where the NBrS cross section is expressed via electron elastic cross section ($\sigma _{el} (E)$). The two terms in the brackets reflect two ways to emit a bremsstrahlung photon, namely after and before a collision. The equation is obtained in the first Born approximation and when one of the partial waves predominates over the others, resulting in that the contribution of interference term is small \cite{Kasyanov65,Kasyanov78}. 

It should be remarked that there are two opinions about what elastic cross section should appear in the formula: that of integral elastic \cite{Firsov61,Kasyanov65,Biberman67,Kasyanov78} or momentum transfer (transport) \cite{Dalgarno66,Biberman67,Park00}. In what follows, we use the integral elastic cross section in Eq. \ref{Eq-sigma-el}.    

Given the differential cross section for neutral bremsstrahlung in elastic collisions (Eq. \ref{Eq-sigma-el}), the spectral emission intensity of photons ($dI_{ph}(\lambda)/d\lambda$) and that of their energy ($dI_{en}(\lambda)/d\lambda$), per unit wavelength and per drifting electron, can be expressed as follows \cite{Kasyanov78,Park00}:
\begin{eqnarray}
\label{Eq-int-ph} 
\frac{dI_{ph}(\lambda)}{d\lambda} = \frac{dN_{ph}}{dt \ N_e \ dV \ d\lambda} = N  \int\limits_{h\nu}^{\infty}\upsilon_e 
\frac{d\sigma}{d\nu} \frac{d\nu}{d\lambda} f(E) \ dE \nonumber \\ 
\mathrm{in \ photon/(s \ nm \ electron)} \; , \\
\label{Eq-int-en}
\frac{dI_{en}(\lambda)}{d\lambda} = N h\nu \int\limits_{h\nu}^{\infty}\upsilon_e  
\frac{d\sigma}{d\nu} \frac{d\nu}{d\lambda} f(E) \ dE \nonumber \\
\mathrm{in \ W/(nm \ electron)} \; ,
\end{eqnarray}
where $N_e$ and $N$ is the electron and atomic density, $dV$ is the volume, $q_e=N_edV$ is the number of drifting electrons, $\upsilon_e=\sqrt{2E/m_e}$ is the electron velocity, $d\sigma/d\nu=(d\sigma/d\nu)_{NBrS,el}$,  $d\nu/d\lambda=-c/\lambda^2$, $f(E)$ is the electron energy distribution function normalized as
\begin{eqnarray}
\label{Eq-norm-f} 
\int\limits_{0}^{\infty} f(E) \ dE = 1 \; .
\end{eqnarray}
For inelastic collisions the equations are the same, except that $d\sigma/d\nu=(d\sigma/d\nu)_{NBrS,exc}$ and the lower limit in the integral is changed to $h\nu + E_{exc}$.

Proportional electroluminescence is characterized by a "reduced EL yield" ($Y_{EL}/N$), which by definition is the number of photons produced per drifting electron per atomic density and per unit drift path. Accordingly, the NBrS contribution to proportional electroluminescence can be described by the following equation (given $dt=dx/\upsilon_d$ in Eq.~\ref{Eq-int-ph}):
\begin{eqnarray}
\label{Eq-NBrS-el-yield} 
\left( \frac{Y_{EL}}{N}\right)_{NBrS} = \frac{dN_{ph}}{dx \ N \ N_e \ dV} = \frac{1}{\upsilon_d \ N} \int\limits_{\lambda_1}^{\lambda_2} \frac{dI_{ph}(\lambda)}{d\lambda} \ d\lambda = \nonumber \\
= \int\limits_{\lambda_1}^{\lambda_2}  \int\limits_{h\nu}^{\infty}\frac{\upsilon_e}{\upsilon_d} 
\frac{d\sigma}{d\nu} \frac{d\nu}{d\lambda} f(E) \ dE \ d\lambda \nonumber \\ 
\mathrm{in \ (photon \ cm^2)/(electron \ atom)}
\; , 
\end{eqnarray}
where $\upsilon_d$ is the electron drift velocity, $\lambda_1-\lambda_2$ is the sensitivity region of the photon detector, $Y_{EL}/N$ is a function of the "reduced electric field" ($\mathcal{E}/N$).

Consequently, the spectrum of the reduced EL yield is 
\begin{eqnarray}
\label{Eq-NBrS-el-yield-spectrum} 
\frac{d (Y_{EL}/N)_{NBrS}}{d\lambda} = 
\int\limits_{h\nu}^{\infty}\frac{\upsilon_e}{\upsilon_d} 
\frac{d\sigma}{d\nu} \frac{d\nu}{d\lambda} f(E) \ dE \  \nonumber \\ 
\mathrm{in \ (photon \ cm^2)/(electron \ atom \ nm)}
\; . 
\end{eqnarray}

The EL yield in this approach is rather sensitive to the electron energy distribution functions. To verify the choice of the distribution function, we applied a similar approach to the ordinary EL mechanism, involving excimers \cite{Chepel13,ArXeN2Proc17,ArELTheory11}:
\begin{eqnarray}
\label{Rea-ord-el}
e^- + Ar \rightarrow e^- + Ar^{\ast} \; , \nonumber \\
Ar^{\ast} + 2Ar \rightarrow Ar^{\ast}_2 + Ar \; , \nonumber \\
Ar^{\ast}_2 \rightarrow 2Ar + h\nu \; .
\end{eqnarray}
Then
\begin{eqnarray}
\label{Eq-ord-el-yield}
\left( \frac{Y_{EL}}{N}\right)_{excimer} =  \int\limits_{E_{exc}}^{\infty}\frac{\upsilon_e}{\upsilon_d} \sigma_{exc}(E) f(E) \ dE 
\; , 
\end{eqnarray}
where $\sigma_{exc}(E)$ is the inelastic cross section to produce an argon excited state in electron-atom collisions. Similarly to \cite{ArELTheory11}, it is assumed here that one excited state (Ar$^{\ast}$) produces one excimer state (Ar$^{\ast}_2$) and that one excimer produces one VUV photon (around 128 nm). Here ordinary electroluminescence is used as a benchmark: if its EL yield calculated this way is consistent with the theoretical and experimental data given elsewhere \cite{ArELTheory11}, then the distribution function can be considered valid. 

\begin{figure}[!htb]
	\centering
	\includegraphics[width=0.99\columnwidth,keepaspectratio]{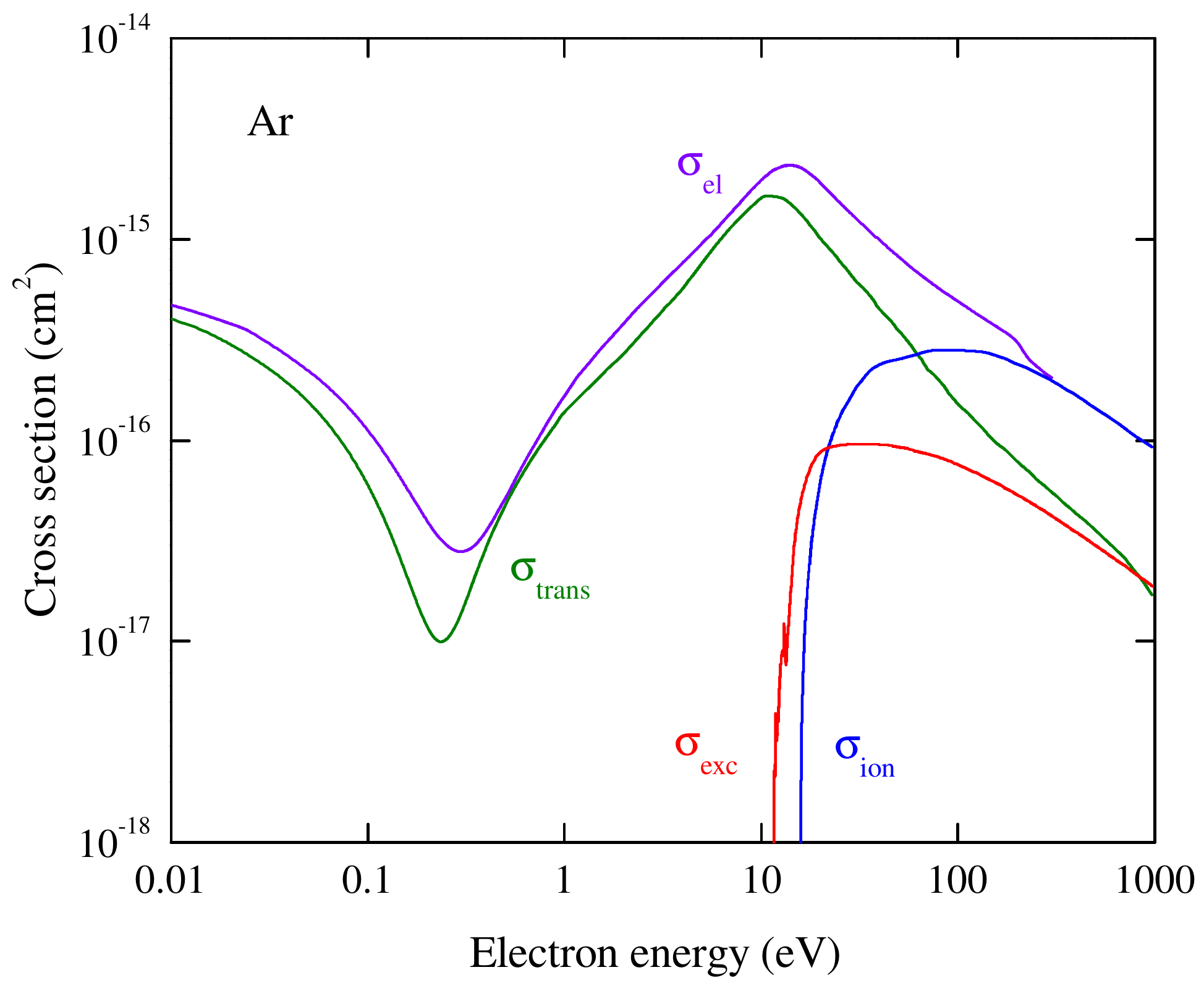}
	\caption{Electron scattering cross sections in argon used in the present work, obtained from the last version of Magboltz 8.97 \cite{Magboltz}: elastic ($\sigma_{el}$), momentum-transfer ($\sigma_{trans}$), excitation ($\sigma_{exc}$) and ionization ($\sigma_{ion}$). The excitation cross section was taken as the sum of those for $>$40 low-lying argon excitation state.}
	\label{CrossSection}
\end{figure}

\section{Cross sections, electron energy distribution functions, NBrS spectra and EL yields}

The electron energy distribution functions were calculated by solving the Boltzmann equation using BOLSIG+ free software \cite{Bolsig1,Bolsig2}, based on electron scattering cross sections from Magboltz \cite{Biagi99,Magboltz}. Fig.~\ref{CrossSection} presents the cross sections used in the present work, namely that of elastic, momentum-transfer, excitation and ionization. The excitation cross section was taken as the sum of those for $>$40 low-lying argon excitation states, the major contribution being provided by the 4 lowest states of Ar$^{\ast}(3p^54s^1)$ configuration.

\begin{figure}[!htb]
	\centering
	\includegraphics[width=0.99\columnwidth,keepaspectratio]{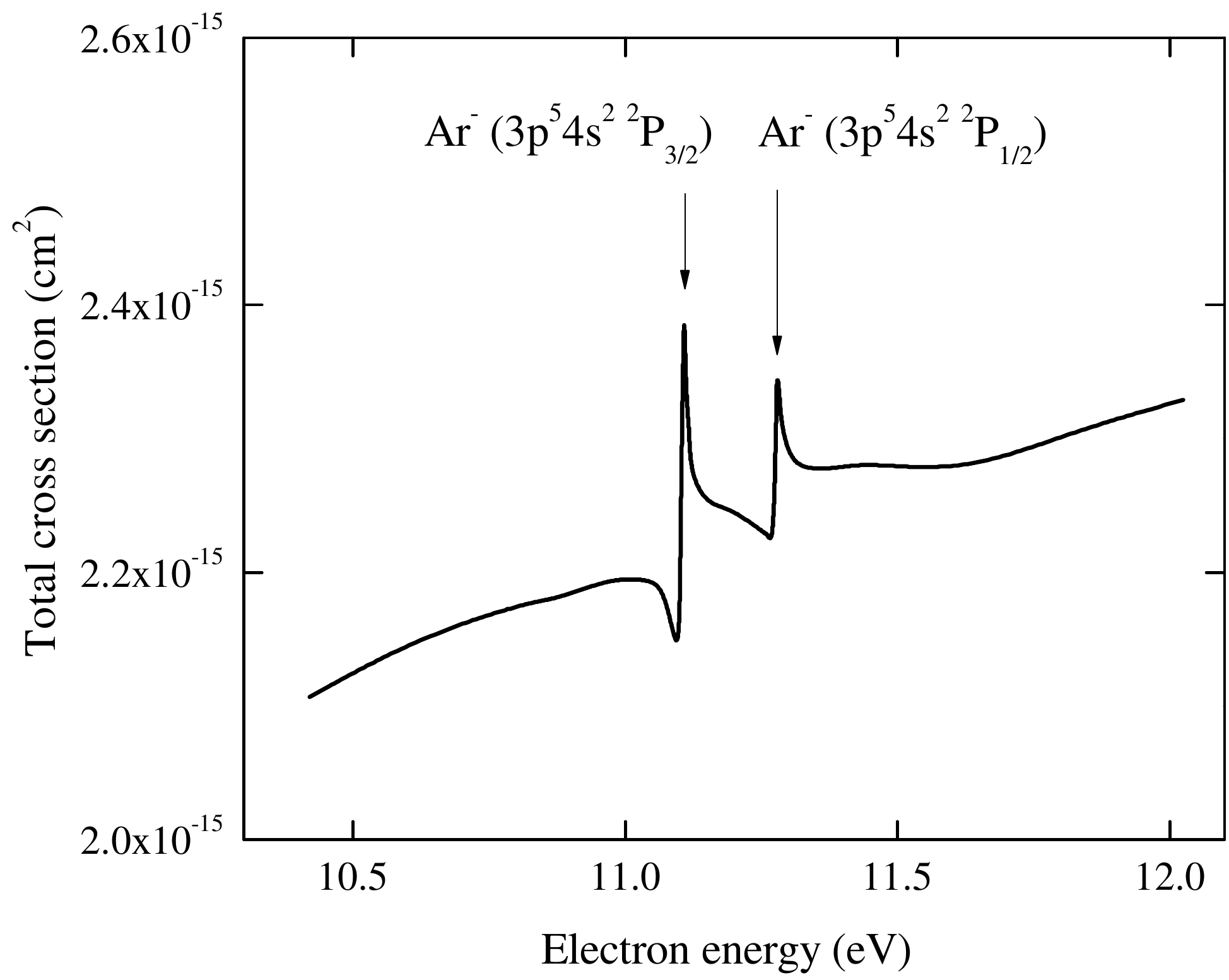}
	\caption{Experimental cross section for electron scattering from Ar around Feshbach resonances \cite{Kurokawa11}.} 
	\label{CrossSection-res}
\end{figure}

Fig.~\ref{CrossSection-res} shows the total cross section measured in \cite{Kurokawa11} around Feshbach resonances \cite{Schulz73}, at 11.10 eV and 11.28 eV, i.e. slightly below the Ar excitation threshold (11.55 eV). The resonance width is $\Gamma$=2-3 meV \cite{Kurokawa11}. Such sub-excitation resonances appear in electron-atom scattering when the electron and the atom form the negative ion bound state:
\begin{eqnarray}
\label{Rea-res}
e^- + Ar \rightarrow Ar^-(3p^54s^2) \rightarrow e^- + Ar \; .
\end{eqnarray}
The cross-section enhancement around the resonances can hardly directly affect the EL yield, since its contribution to the integral in Eq. \ref{Eq-NBrS-el-yield}  is not that large. However it was noted that the resonances may have an indirect effect: at the resonance energies the electron energy distribution function might be modified \cite{DeMunari71,DeMunari84} and the NBrS intensity might be enhanced \cite{Dyachkov74,Dallacasa80}. The latter issue will be  discussed further at the end of the section.

\begin{figure}[!htb]
	\centering
	\includegraphics[width=0.99\columnwidth,keepaspectratio]{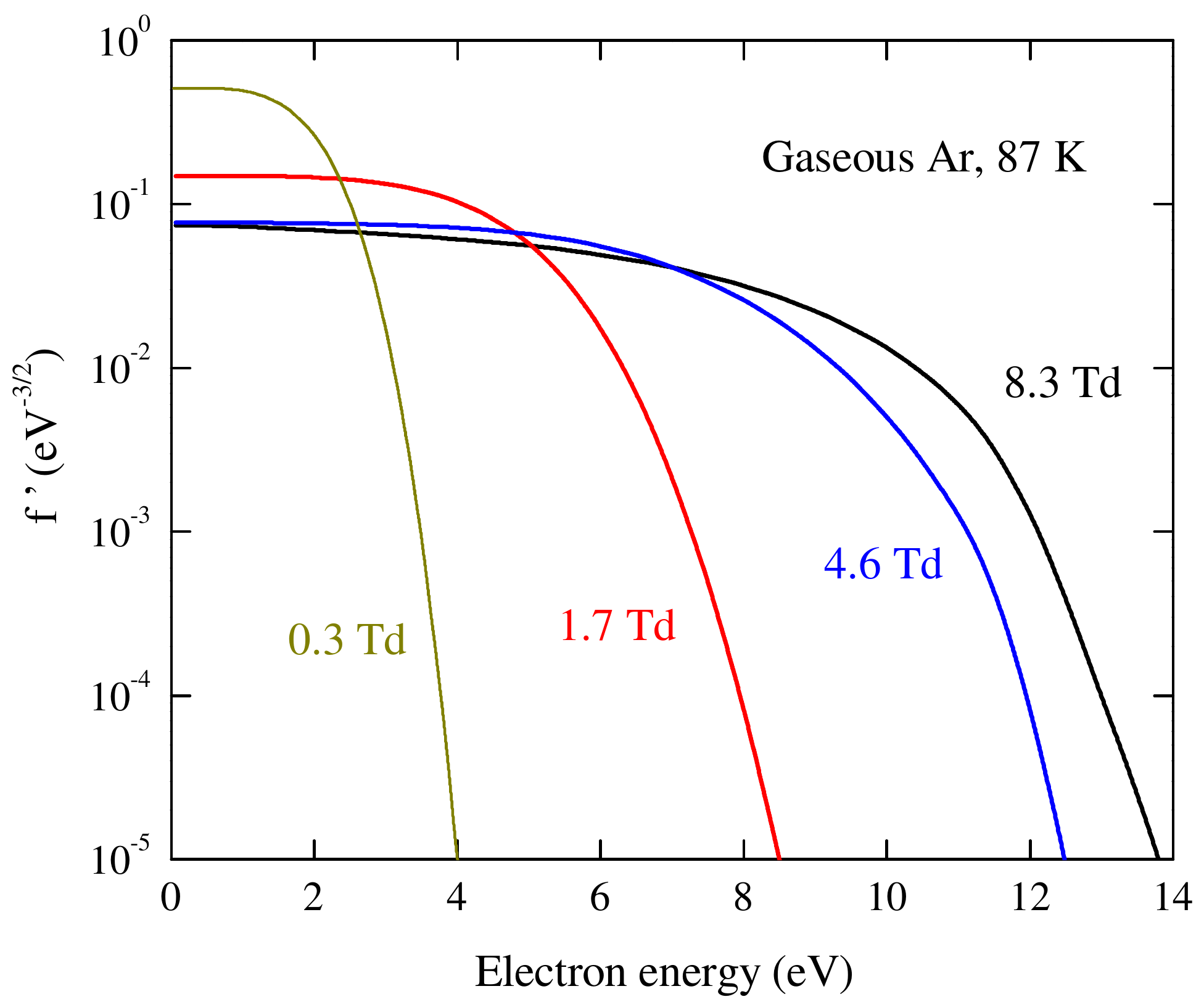}
	\caption{Electron energy distribution functions with a prime, normalized as Eq.~\ref{Eq-norm-fprime}, calculated using Boltzmann equation solver BOLSIG+  \cite{Bolsig2}, at different reduced electric fields.}
	\label{DistFunction}
\end{figure}

\begin{figure}[!htb]
	\centering
	\includegraphics[width=0.99\columnwidth,keepaspectratio]{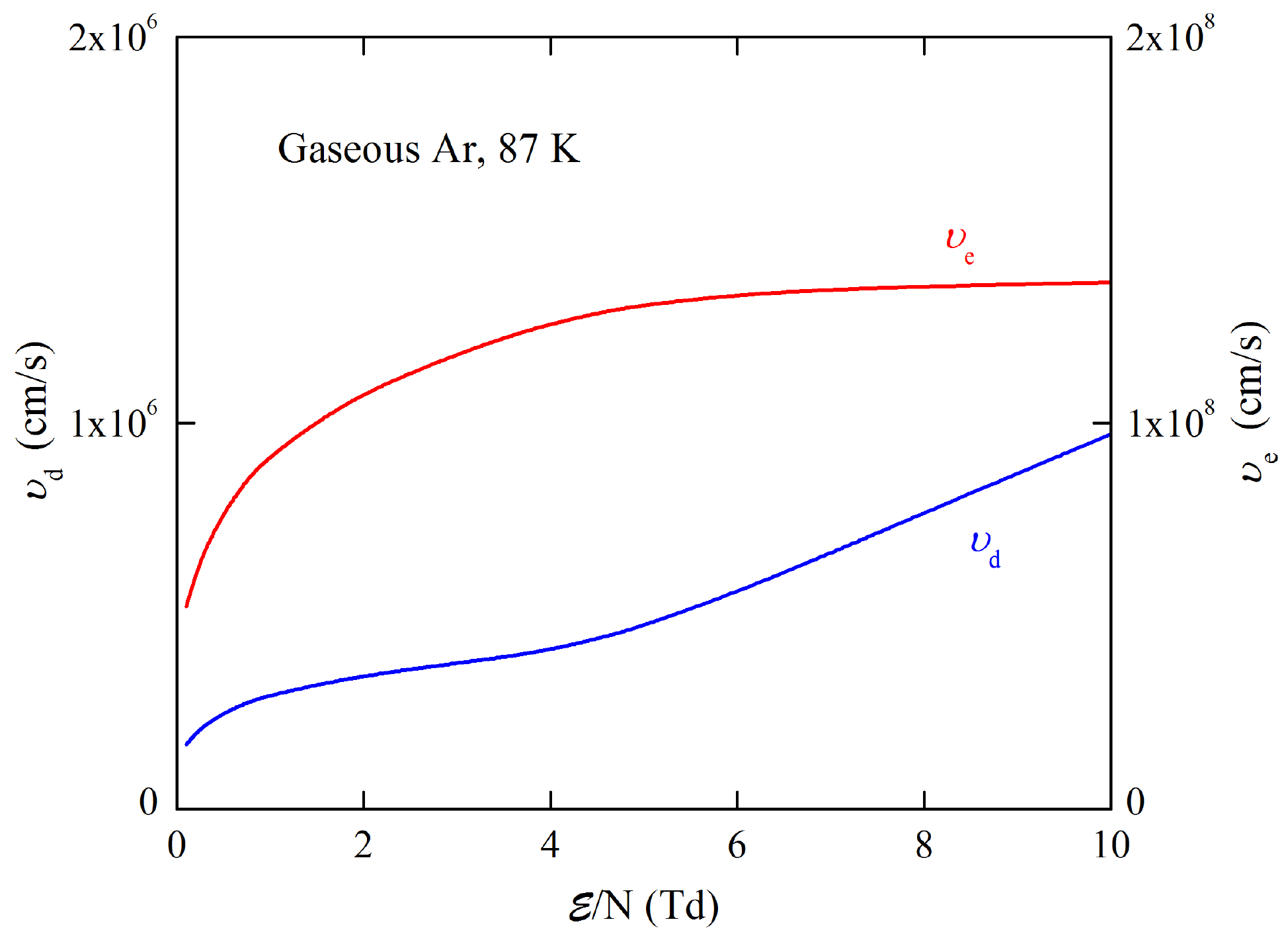}
	\caption{Average velocity of directed motion of electrons (drift velocity, $\upsilon_d$) and that of chaotic motion ($\upsilon_e$) as a function of the reduced electric field, calculated using Boltzmann equation solver.}
	\label{Velocity}
\end{figure}

\begin{figure}[!htb]
	\centering
	\includegraphics[width=0.99\columnwidth,keepaspectratio]{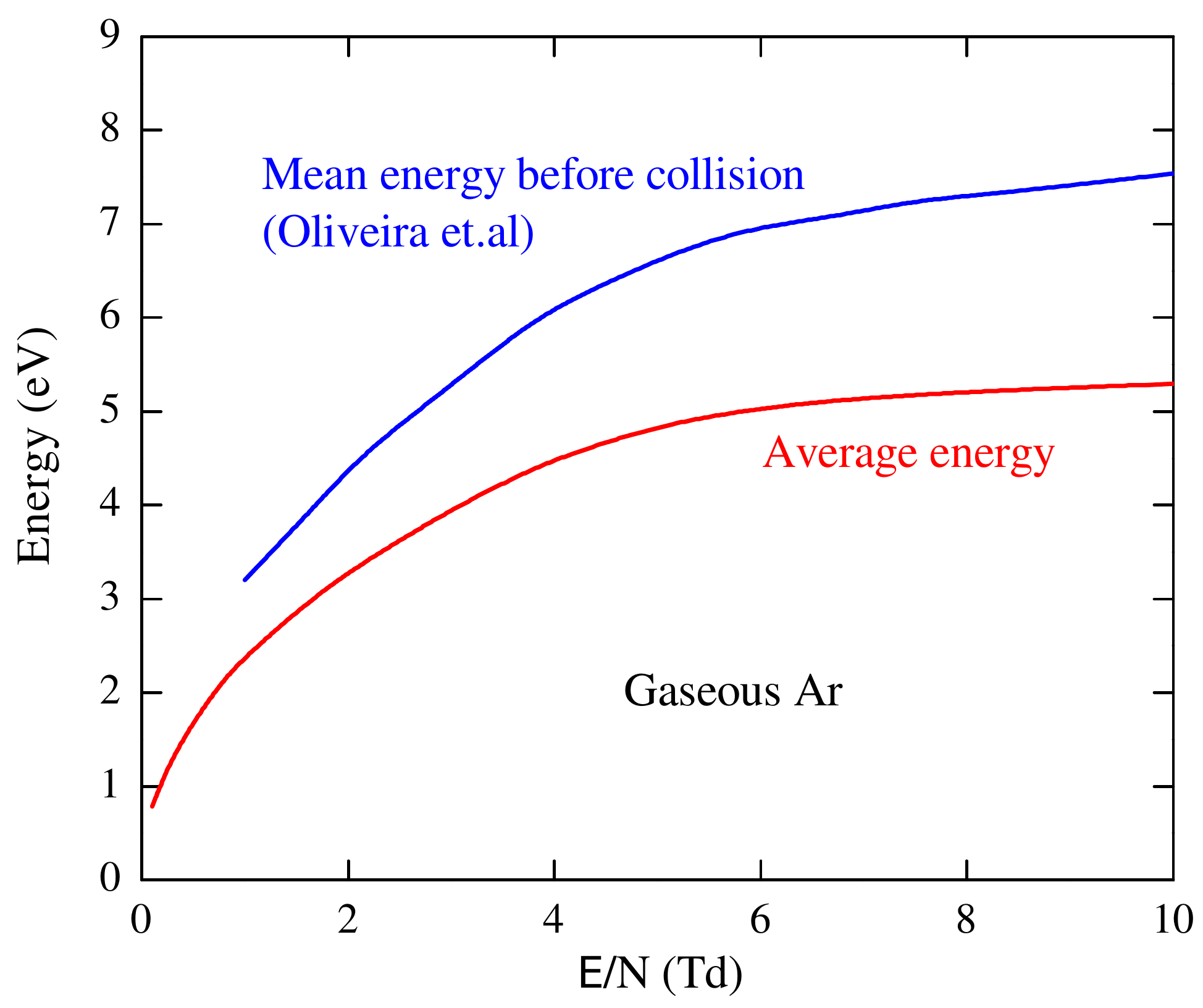}
	\caption{Average electron energy as a function of the reduced electric field, calculated using Boltzmann equation solver. Also shown is the mean energy reached by the electrons before collisions, taken from Oliveira et al. \cite{ArELTheory11}.}
	\label{Energy}
\end{figure}

Fig.~\ref{DistFunction} shows examples of the calculated electron energy distribution functions, namely the distribution functions with a prime ($f^\prime$) often used instead of $f$ (see Eq.~\ref{Eq-norm-f}) and normalized as 
\begin{eqnarray}
\label{Eq-norm-fprime} 
\int\limits_{0}^{\infty} E^{1/2} f^\prime(E) \ dE = 1 \; .
\end{eqnarray}
$f^\prime$ is considered to be more enlightening
than $f$ since in the limit of zero electric field it tends to Maxwellian distribution. Reduced electric fields ($\mathcal{E}/N$) in the figure are expressed in Townsends (1 Td = 10$^{-17}$ V cm$^2$); the field range includes that used in our measurements, namely that varying from 1.7 to 8.3 Td, corresponding to 1.4 and 7.2 kV/cm (in gaseous Ar at 87 K and 1.0 atm). Note that the reduced field of 4.6 Td corresponds to the electric field of 4.0 kV/cm, which is equivalent to that used in the DarkSide-50 dark matters search experiment \cite{DarkSide15}.  

Figs.~\ref{Velocity} and \ref{Energy} show the quantities calculated along with the distribution function: the average velocity of directed motion of electrons (drift velocity, $\upsilon_d$), that of chaotic motion ($\upsilon_e$) and the average electron energy.

\begin{figure}[!htb]
	\centering
	\includegraphics[width=0.99\columnwidth,keepaspectratio]{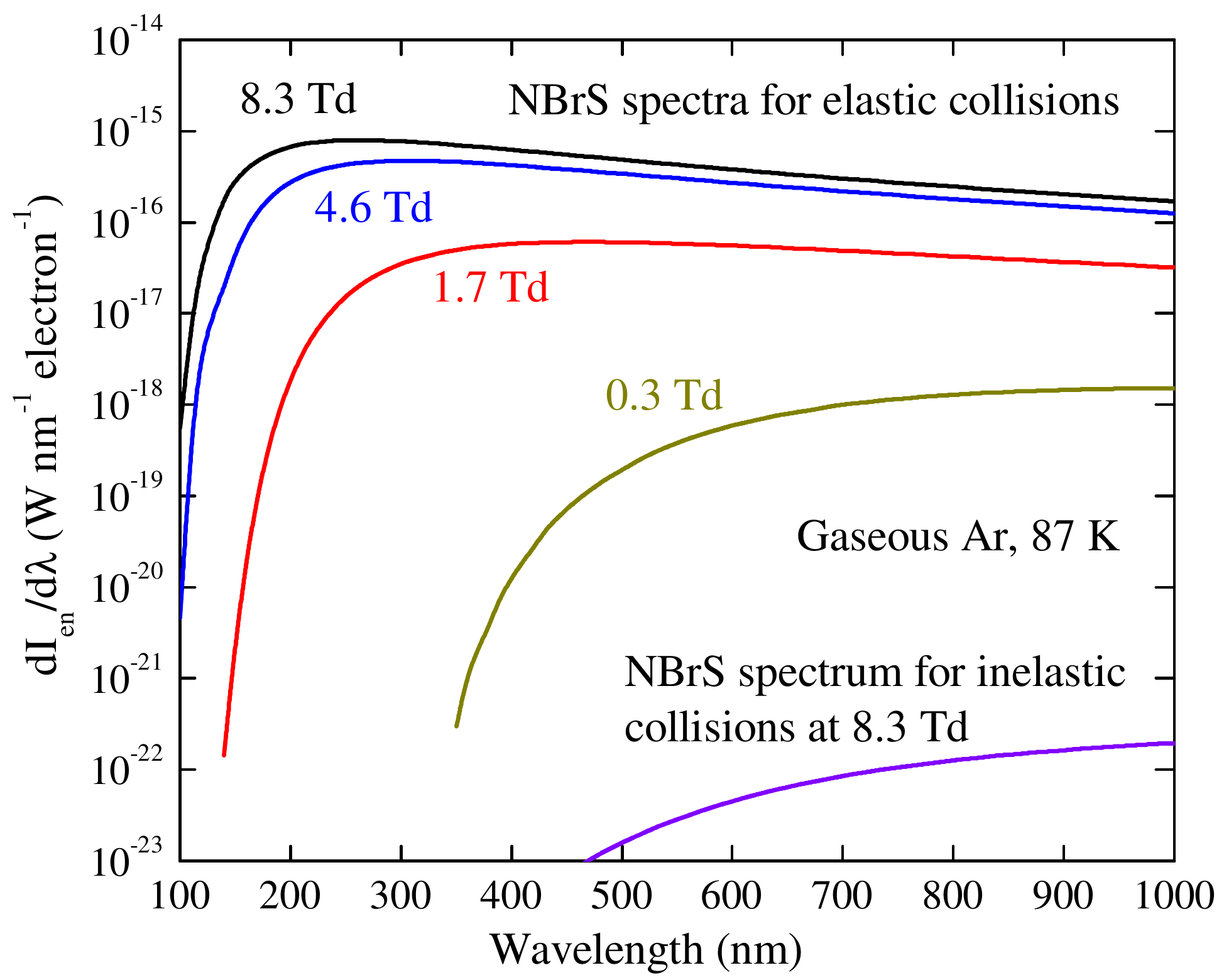}
	\caption{Spectra of NBrS energy intensity for proportional electroluminescence in gaseous Ar at different reduced electric fields, calculated for elastic collisions using Eq.~\ref{Eq-int-en}, where the distribution functions were obtained using Boltzmann equation solver. In addition, a NBrS spectrum for inelastic collisions at 8.3 Td is shown. 
	}
	\label{NBrS-spectra-int}
\end{figure}

\begin{figure}[!htb]
	\centering
	\includegraphics[width=0.99\columnwidth,keepaspectratio]{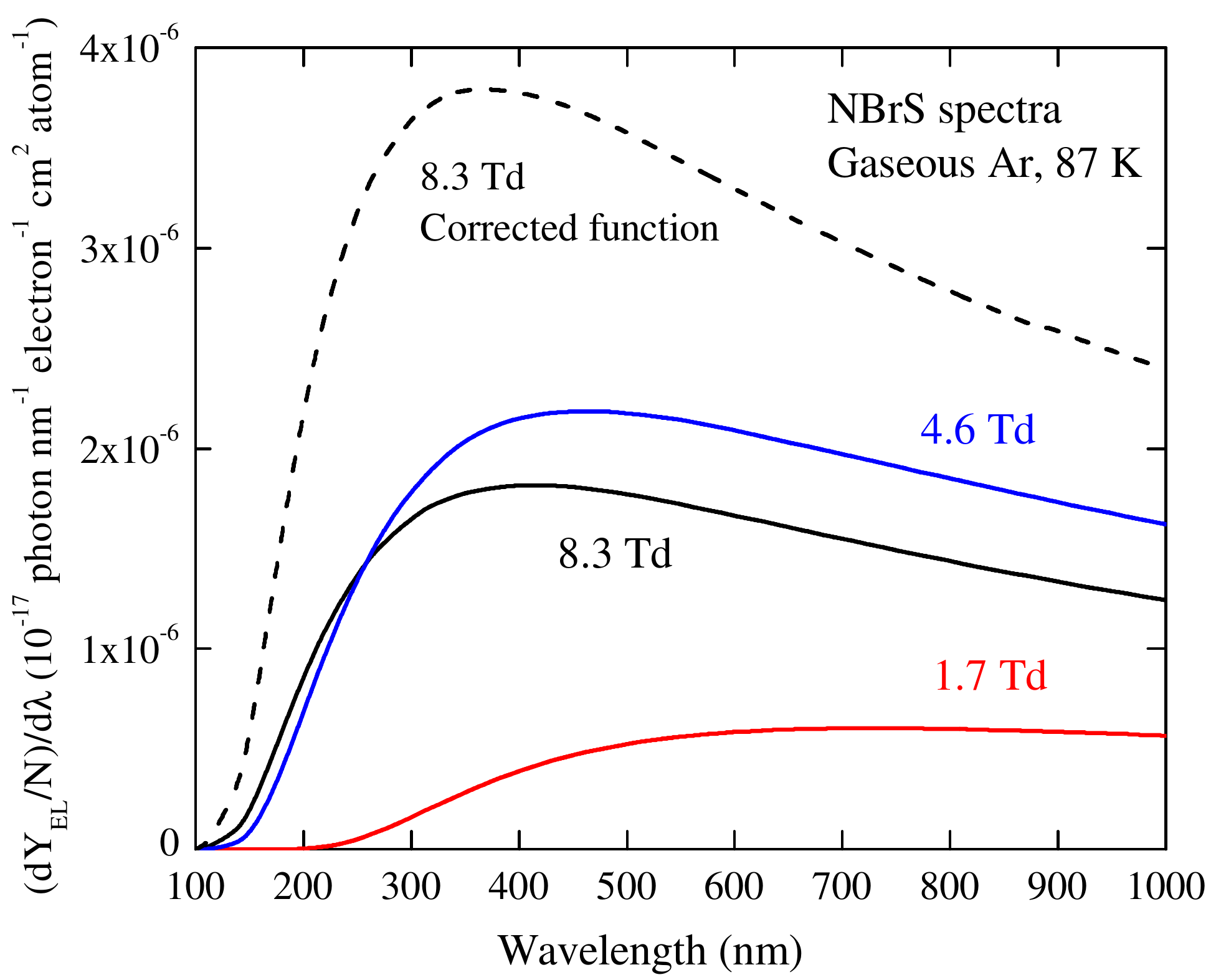}
	\caption{Spectra of NBrS reduced EL yield in gaseous Ar at different reduced electric fields, calculated using Eq.~\ref{Eq-NBrS-el-yield-spectrum}, where the distribution functions were obtained using Boltzmann equation solver (solid lines). Also shown is the spectrum at 8.4 Td obtained using the distribution function corrected for mean energy before collisions (dashed line, "corrected function").}
	\label{NBrS-EL-yield-spectra}
\end{figure}

\begin{figure}[!htb]
	\centering
	\includegraphics[width=0.99\columnwidth,keepaspectratio]{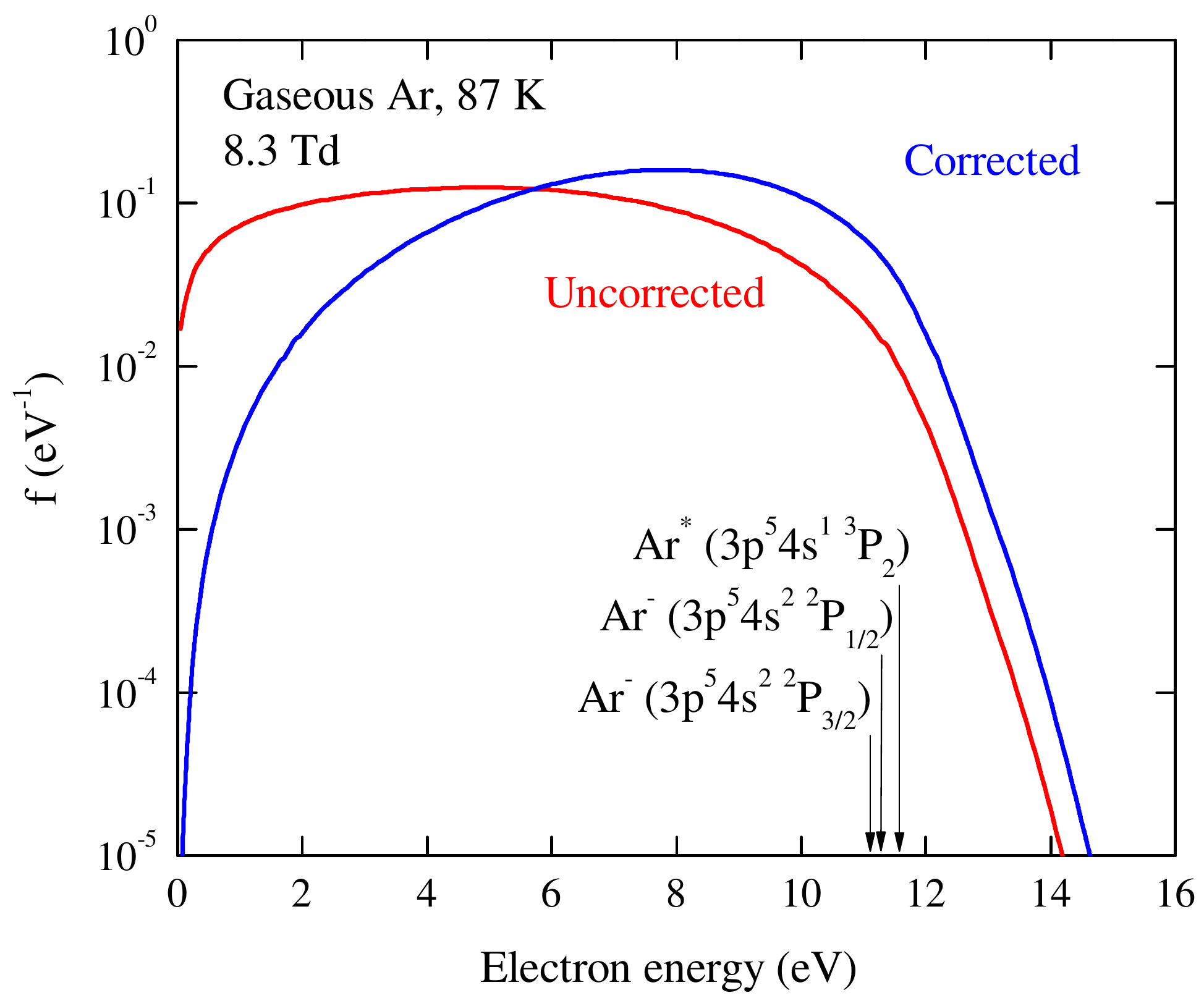}
	\caption{Electron energy distribution function at 8.3 Td, normalized as Eq.~\ref{Eq-norm-f}, obtained using Boltzmann equation solver, before and after correction for mean energy before collisions. The arrows indicate the positions of Feshbach resonances and that of the lowest argon excitation level.}
	\label{DistFunctionCorr}
\end{figure}

Figs.~\ref{NBrS-spectra-int} and \ref{NBrS-EL-yield-spectra} show the NBrS spectra of the photon energy intensity and reduced EL yield.  The spectra were calculated by numerical integration in Eq.~\ref{Eq-int-en} and Eq.~\ref{Eq-NBrS-el-yield-spectrum}. One can see that the spectra are rather flat, extending from the UV to visible and NIR regions. The spectra looks very similar to those calculated for NBrS in glow discharge \cite{Rutscher76}, that speaks in favor of correctness of calculations.
In addition, a NBrS spectrum for inelastic collisions at the highest electric field is shown in Fig.~\ref{NBrS-spectra-int}, to demonstrate that its contribution is considerably (6 orders of magnitude) smaller compared to that of elastic collisions. That is why in what follows we neglect the NBrS contribution due to inelastic collisions. 

\begin{figure}[!htb]
	\centering
	\includegraphics[width=0.99\columnwidth,keepaspectratio]{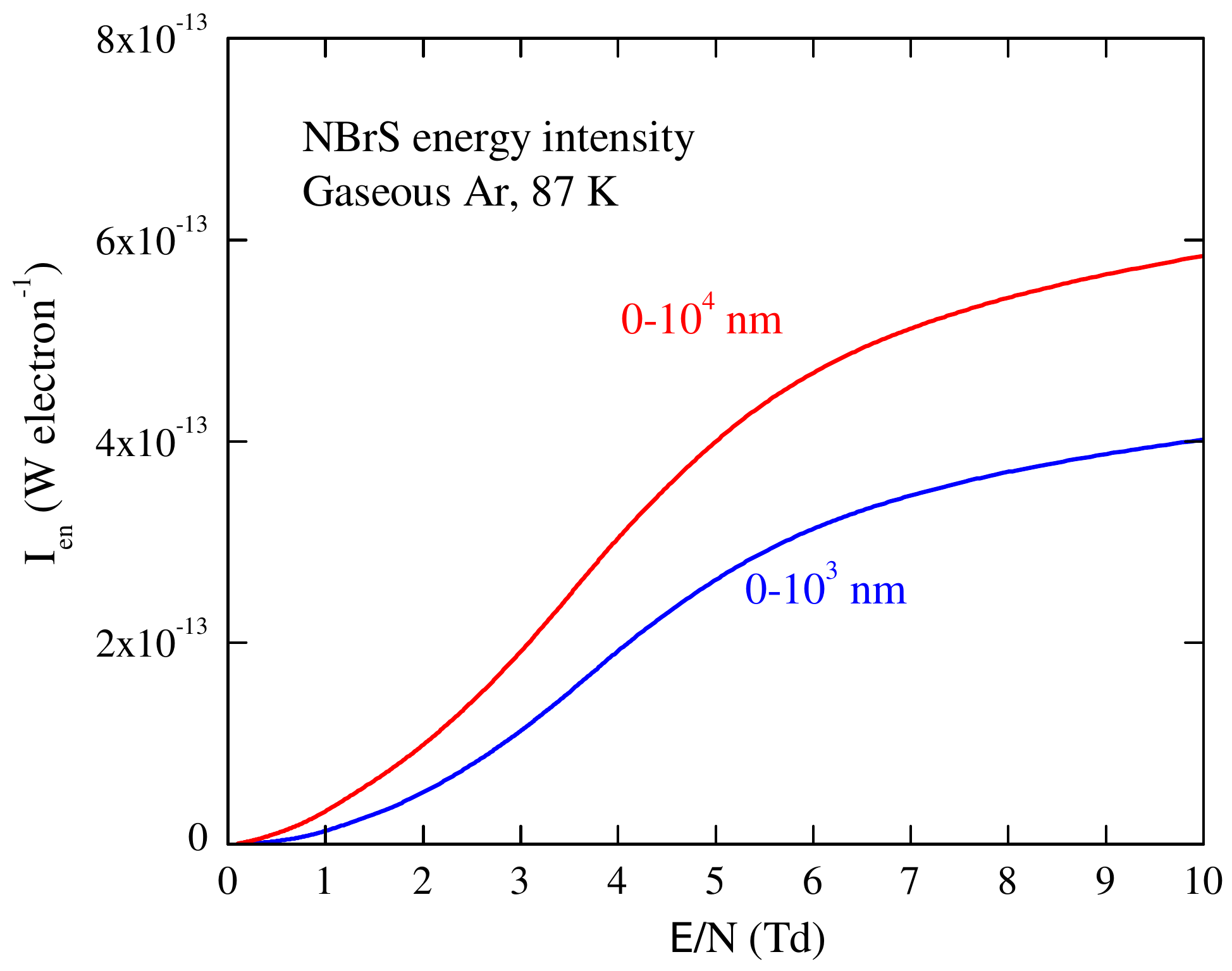}
	\caption{Integral  NBrS energy intensity ($I_{en}$) for proportional electroluminescence in gaseous Ar as a function of the reduced electric field, integrated over two different wavelength regions, that of 0-1000 nm and 0-10000 nm. The quantity was calculated integrating Eq.~\ref{Eq-int-en} where the distribution functions were obtained using Boltzmann equation solver.}
	\label{NBrS-int-field-dep}
\end{figure}

\begin{figure}[!htb]
	\centering
	\includegraphics[width=0.99\columnwidth,keepaspectratio]{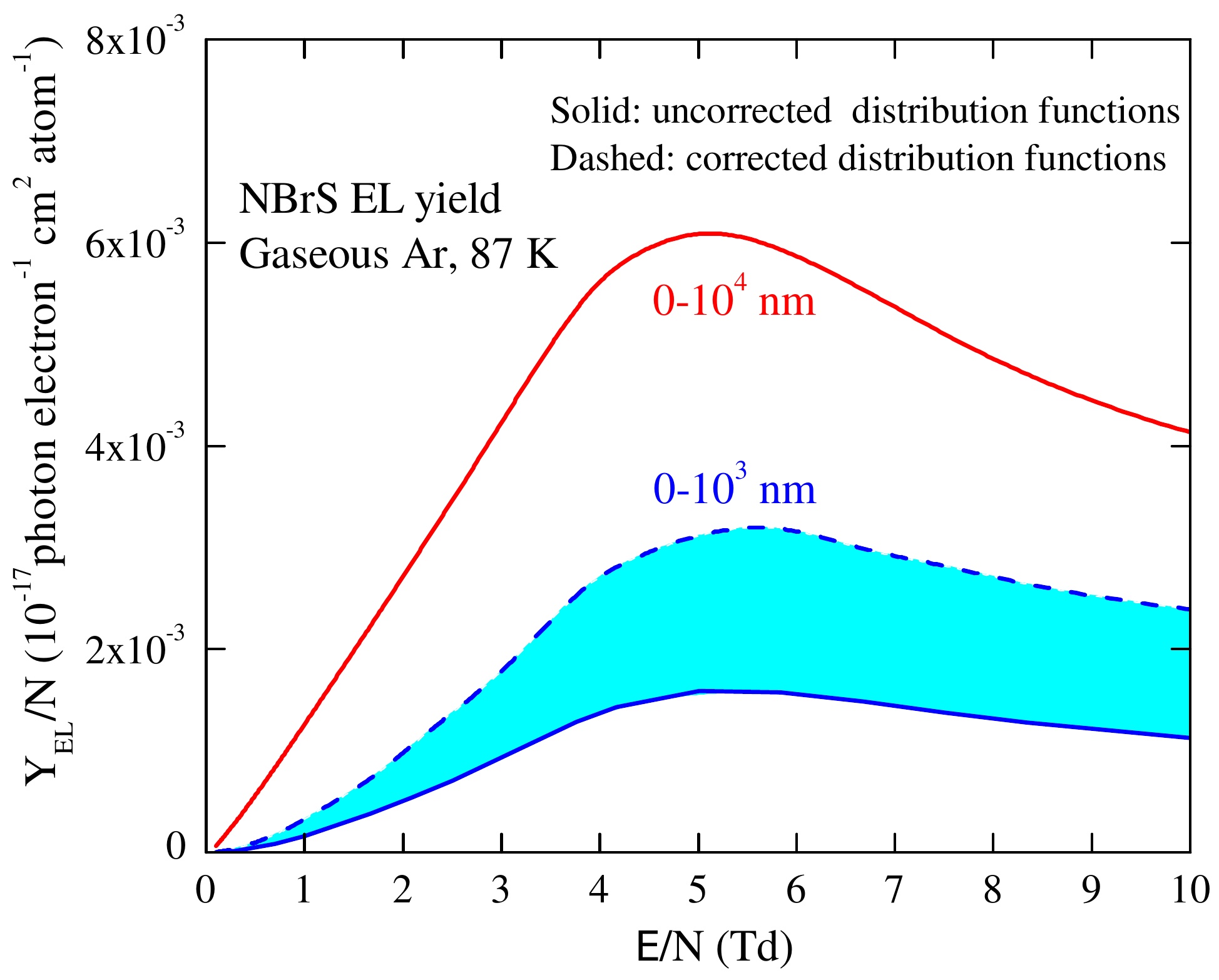}
	\caption{Reduced EL yield ($Y_{EL}/N$) due to neutral bremsstrahlung in gaseous Ar as a function of the reduced electric field, integrated over two different wavelength regions, that of 0-1000 nm and 0-10000 nm, using Eq.~\ref{Eq-NBrS-el-yield}. For 0-1000 nm region it was calculated for two types of distribution functions: those obtained using Boltzmann equation solver (uncorrected distribution functions) and those corrected for mean energy before collisions (corrected distribution functions).}
	\label{NBrS-yield-field-dep}
\end{figure}

\begin{figure}[htb]
	\centering
	\includegraphics[width=0.99\columnwidth,keepaspectratio]{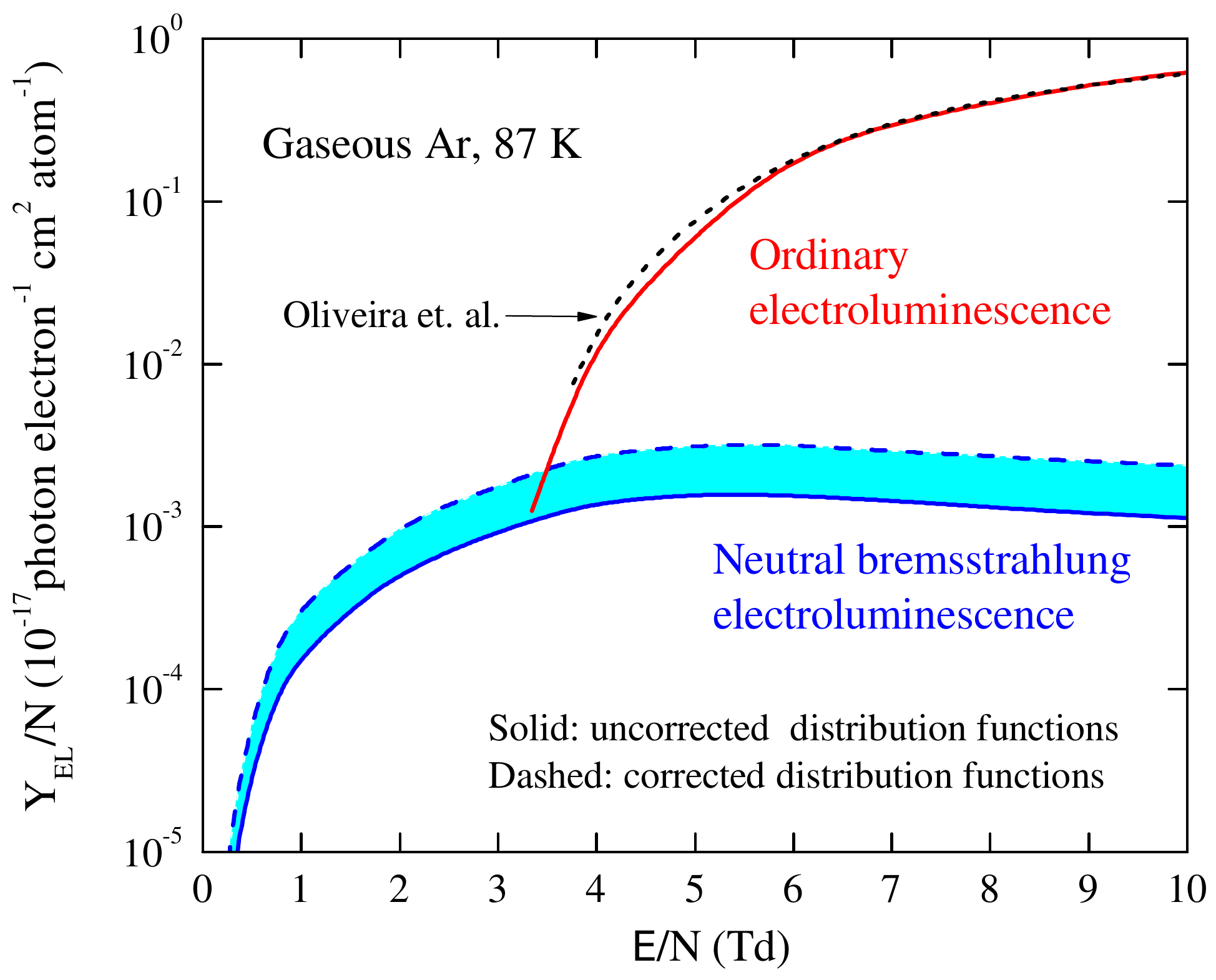}
	\caption{Reduced ordinary EL yield and that of neutral bremsstrahlung at 0-1000 nm in gaseous Ar as a function of the reduced electric field. The quantity was calculated for two types of distribution functions: those obtained using Boltzmann equation solver (uncorrected distribution functions) and those corrected for mean energy before collisions (corrected distribution functions). For comparison, the EL yield calculated using microscopic approach from Oliveira et al. \cite{ArELTheory11} is shown.}
	\label{NBrS-vs-ord-EL-yield}
\end{figure}

It should be remarked that the approach based on the solution of the Boltzmann equation is rather approximate, resulting in that Figs.\ref{NBrS-spectra-int} and \ref{NBrS-EL-yield-spectra} give only a rough idea of the NBrS spectra. Indeed, the distribution function and other quantities in Figs.~\ref{DistFunction},\ref{Velocity} and \ref{Energy} obtained by solving the Boltzmann equation, are actually averaged over the collisions and hence also over the time between the collisions. In fact, the energy in equations \ref{Eq-int-ph},\ref{Eq-int-en},\ref{Eq-NBrS-el-yield} and \ref{Eq-NBrS-el-yield-spectrum} should be the energy just before the collisions, which obviously  exceeds the time-averaged energy since the electrons are accelerated by the field between the collisions. This difference is illustrated in Figs.~\ref{Energy} where the average electron energy, obtained solving Boltzmann equation, is compared to the mean energy before collisions, obtained in \cite{ArELTheory11} by simulation of electron transport in an electric field using microscopic approach. 

We tried to estimate the effect of this difference on the EL yields: the distribution functions were corrected by multiplying those by a power function so that the average energies for the corrected distributions coincide with the mean energies before collisions taken from \cite{ArELTheory11}. An example of this correction procedure is shown in Fig.~\ref{DistFunctionCorr}. 

Figs.~\ref{NBrS-int-field-dep} and \ref{NBrS-yield-field-dep} show the integral  NBrS photon energy intensity ($I_{en}$) and reduced EL yield ($Y_{EL}/N$) as a function of the reduced electric field, integrated over two different wavelength regions: that of 0-1000 nm and 0-10000 nm. In addition, Fig.~\ref{NBrS-vs-ord-EL-yield} compares the ordinary EL yield, calculated using Eq.~\ref{Eq-ord-el-yield}, to that of neutral bremsstrahlung, calculated using Eq.~\ref{Eq-NBrS-el-yield}.  One can see that ordinary electroluminescence has a characteristic field dependence: it starts at some threshold electric field, of about of 4.0 Td, corresponding to the commencement of atomic excitations, and then linearly increases with the field.

The NBrS yield in the latter two figures was calculated for the two considered types of the distribution functions: obtained using Boltzmann equation solver and corrected for mean energy before collisions. The effect of the corrected functions resulted in an increase of the EL yield by a factor of 2 for NBrS electroluminescence, indicating that the prediction uncertainty of the theory developed here is not that small. Within this uncertainty, as one can see from Fig.~\ref{NBrS-vs-ord-EL-yield}, the ordinary electroluminescence yield calculated in this work is consistent with that of using microscopic approach \cite{ArELTheory11}.

The NBrS EL yield in the 0-1000 nm range is of practical importance, since its long-wavelength limit is defined by SiPM sensitivity, thereby representing the maximum number of NBrS photons that can ever be detected by existing devices. The NBrS EL yield in the 0-10000 nm range is of hypothetical interest: it demonstrates  that the NBrS signal could be considerably increased, by a factor of 5, if one had a single-photon detector in the infrared. 

One can see from Fig.~\ref{NBrS-yield-field-dep} that the NBrS EL yield first increases with the field, then saturates and even decreases at some field values (in contrast to photon energy intensity that increases monotonically with the field). Such a behavior, at a slowly varying elastic cross section (shown in Fig. \ref{CrossSection}), reflects that of the ${\upsilon_e}/{\upsilon_d}$ ratio: see Fig.~\ref{Velocity} and Eq.~\ref{Eq-NBrS-el-yield}.

Finally, let us discuss the enhancement of NBrS emission on resonance, considered above. One of the suggested mechanisms of the enhancement is that the electron attachment on the resonance level and the subsequent decay of the negative ion are accompanied by photon emission \cite{Dyachkov74}:  
\begin{eqnarray}
\label{NBrS-Res}
e^- + Ar \rightarrow Ar^- + h\nu \; ; \\ Ar^- \rightarrow e^- + Ar + h\nu \; .
\end{eqnarray}
In \cite{Dyachkov74} these processes were calculated for atomic nitrogen using the electron radial wave functions on the resonance (extrapolated from the other state): it was shown that the NBrS effect on N$^-$ resonance may enhance the overall NBrS intensity by an order of magnitude. It is obvious that for each particular resonance the degree of such enhancement is different. Unfortunately for Ar$^-$ resonance these radial wave functions are unknown; accordingly, the calculations cannot be performed at the moment. It should be remarked that since the resonance levels (11.10 eV and 11.28 eV) are very close to those of excitation (11.55 eV), the effect of neutral bremsstrahlung electroluminescence on resonance would have the same threshold for the electric field as that of ordinary electroluminescence, i.e. around 4.0 Td. Below this threshold, the NBrS effect on resonance would not come into force and thus the predictions of the theory would not change. 

Summarizing, the theory of neutral bremsstrahlung predicts the following properties of proportional electroluminescence in two-phase Ar: \\
1)  electroluminescence below the Ar excitation threshold, in the UV, visible and NIR regions; \\
2) appreciable non-VUV component above the Ar excitation threshold, extending from the UV to NIR.

In the following sections, we will see how much these predictions of the theory are confirmed in experiment.

\section{Experimental setup and procedures}

\begin{figure}[htb]
	\centering
	\includegraphics[width=0.99\columnwidth,keepaspectratio]{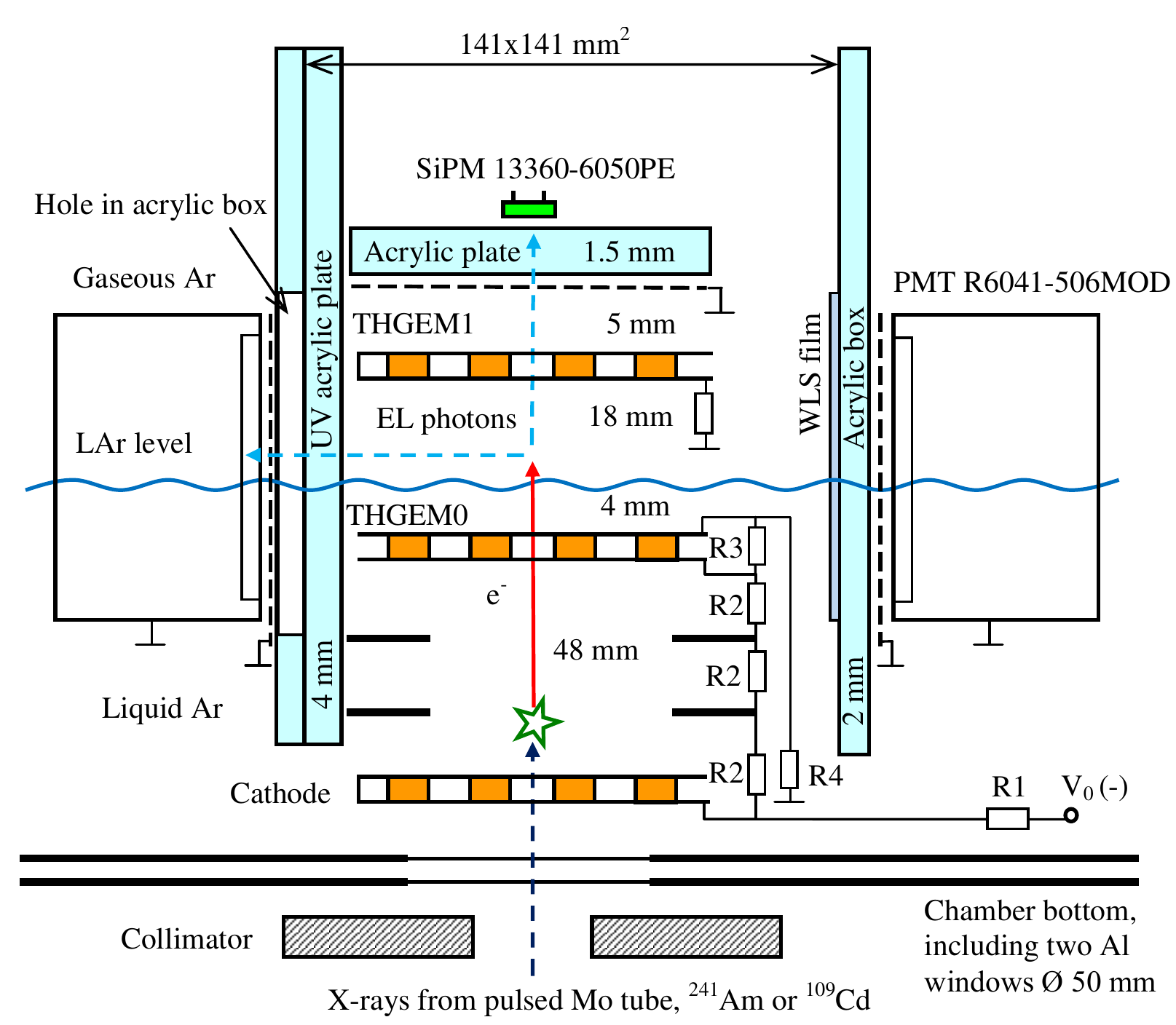}
	\caption{Schematic view of the experimental setup (not to scale).}
	\label{Setup}
\end{figure}

The experimental setup was similar to that used in our previous measurements \cite{CRADPropEL17}: see Fig.~\ref{Setup}. The setup included a two-phase detector with EL gap, filled with 2.5 liters of liquid Ar. In contrast to previous measurements where Ar was doped with N$_2$ \cite{CRADPropEL15,CRADELGap17,CRADPropEL17}, in this work we used pure Ar. Argon, of an initial purity of 99.9998\%, was used on a closed loop; it was purified from electronegative impurities by Oxisorb filter each time during the cooling procedure, providing electron life-time in the liquid $>$100 $\mu$s \cite{CRADELGap17}. The N$_2$ content was below 1 ppm (with the measurement accuracy of 1 ppm); it was constantly monitored before and after each cryogenic run using gas analyzer "SVET" \cite{SVET}, which employed emission-spectrum-measurement technique. The detector was operated in a two-phase mode in the equilibrium state, at a saturated vapor pressure of 1.000$\pm$0.005 atm and at a temperature of 87.3 K. 

The EL gap, formed by the liquid surface and a THGEM plate (THGEM1 in the figure), was viewed by four compact 2-inch PMTs R6041-506MOD \cite{CryoPMT15,CryoPMT17}, located on the perimeter of the gap. The PMTs were electrically insulated from the gap by an acrylic box. Three of the four PMTs were made sensitive to the VUV emission of pure Ar by depositing a WLS films (based on TPB in polystyrene matrix \cite{TPB13}) on the inner box surface facing the EL gap in front of these PMTs. Let us designate this readout configuration as 3PMT+WLS. 

An important modification of the setup compared to that of \cite{CRADPropEL15,CRADPropEL17} was that one of the four PMTs  was intentionally left insensitive to the VUV (1PMT readout configuration). Instead, its sensitivity to the UV was provided, starting from 300 nm: in place of the acrylic with the WLS film, a hole was made in the box, which was isolated from the EL gap by an UV-acrylic plate transparent above 300 nm. 

The EL gap was also viewed by a 11x11 SiPM-matrix from the top, through an acrylic plate and a THGEM, with a spectral sensitivity ranging from the near UV (360 nm) to the NIR (1000 nm). Only the central SiPM, of 6x6 mm$^2$ area and MPPC 13360-6050PE type \cite{Hamamatsu}, was used in the present work; it was operated at overvoltage of either 3.6 or 5.6 V. 

Another modification was that we used one THGEM in front of the SiPM matrix, instead of the two in \cite{CRADPropEL15,CRADPropEL17}, for simplicity and to increase the SiPM-matrix response in direct optical readout of the EL gap in future studies. The detector was irradiated from outside by X-rays from a pulsed X-ray tube with the average energy of 25 keV \cite{XRayYield16}, those from $^{109}$Cd source with the energy of 22-25 and 88 keV, and those from $^{241}$Am source with the energy of 60 keV.

The direct outputs from all the PMTs were amplified using fast 10-fold amplifiers CAEN N979; the fast output signals were used in time measurements for pulse-shape analysis. The signals from the 3 PMTs with WLS were summed (using CAEN N625 unit) and re-amplified with a linear amplifier with a shaping time of 200 ns, for the subsequent amplitude analysis. The same was done for the PMT without WLS. The SiPM signals were recorded using dedicated fast amplifiers with a shaping time of 40 ns.

The DAQ system included both a 4-channel oscilloscope LeCroy WR HRO 66Zi and a 64-channel Flash ADC CAEN V1740 (12 bits, 62.5 MHz): the signals were digitized and stored both in the oscilloscope and in a computer for further off-line analysis.

\begin{figure}[htb]
	\centering
	\includegraphics[width=0.99\columnwidth,keepaspectratio]{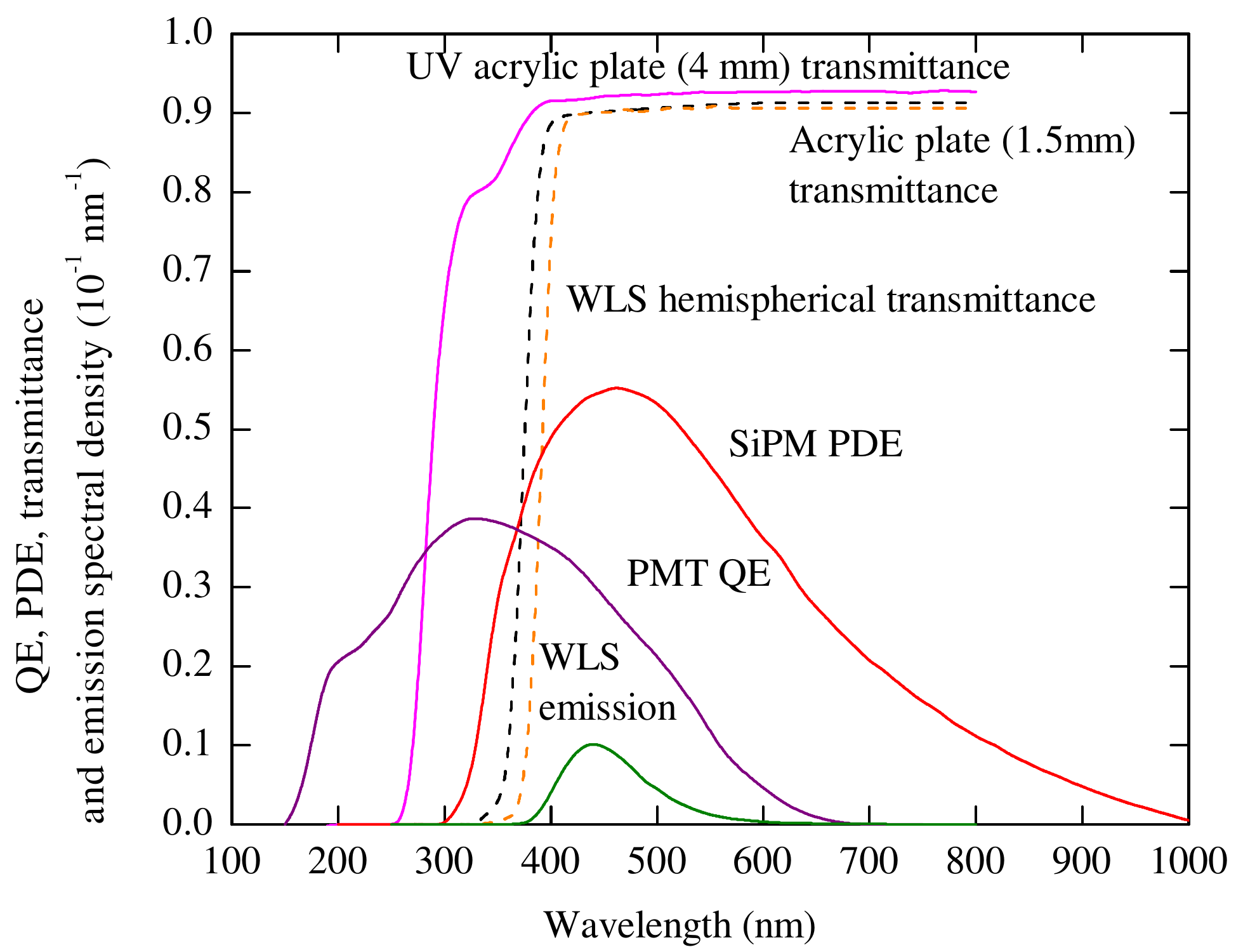}
	\caption{Quantum efficiency (QE) of the PMT R6041-506MOD at 87 K obtained from \cite{Hamamatsu,PMTQE} using a temperature dependence derived there, Photon Detection Efficiency (PDE) of the SiPM (MPPC 13360-6050PE \cite{Hamamatsu}) at an overvoltage of 5.6 V obtained from \cite{SiPM17} using the PDE voltage dependence, transmittance of the ordinary and UV acrylic plate in front of the SiPM and 1PMT, respectively, measured by us, and hemispherical transmittance of the WLS (TPB in polystyrene) \cite{TPB13a}. Also shown is the emission spectrum of the WLS (TPB in polystyrene) \cite{TPB13}.}
	\label{QEPDE}
\end{figure}

Fig.~\ref{QEPDE} presents optical spectra of the PMT Quantum Efficiency (QE) \cite{Hamamatsu,PMTQE}, SiPM Photon Detection Efficiency (PDE) \cite{CryoMPPC15,Hamamatsu,SiPM17}, transmittance of the ordinary and UV acrylic plate in front of the SiPM and 1PMT, respectively, and WLS hemispherical transmittance \cite{TPB13a}. In addition, the emission spectrum of the WLS \cite{TPB13}  is presented. These data were used to determine the absolute EL yield. 

Thus we had three ways of optical readout of the EL gap, with the appropriate photodetector configuration designated here as 1PMT, 3PMT+WLS and SiPM. Such a combination of photodetectors, sensitive in different spectral ranges, made it possible to assess the spectral composition of EL radiation. The appropriate EL gap yield is defined as the number of photoelectrons recorded by any configuration ($N_{pe}$), normalized to the number of drifting electrons in the EL gap ($q_e$):
\begin{equation}
\label{Eq-elgap-yield} 
Y_{XX}=N_{pe}/q_e \ ; \ $XX=1PMT, 3PMT+WLS or SiPM$ \ .
\end{equation}
At higher electric fields, $q_e$ was measured directly at the THGEM1 electrode acting as an anode of the EL gap, using a calibrated circuit of a preamplifier and shaping amplifier. At lower electric fields, $q_e$ was determined by extrapolation using the field dependence of the X-ray ionization yield in liquid Ar \cite{XRayYield16}. At that, the electron transmission through the THGEM0 electrode, determined by us using 3D simulation, and the emission efficiency through the liquid-gas interface, studied elsewhere \cite{EEmissEff},  were taken constant in a given field range. 

The absolute EL yield defined in Eq.~\ref{Eq-NBrS-el-yield} is rewritten here through the quantities measured in experiment:
\begin{equation}
\label{Eq-el-yield} 
Y_{EL}=N_{ph}/q_e/d \,,
\end{equation}
where $d$ is the EL gap thickness and the number of photons recorded by the PMTs or SiPM is defined as
\begin{equation}
\label{Eq-n-ph} 
N_{ph}=N_{pe}/PCE \,.
\end{equation}
Here $PCE$ is the photon-to-photoelectron conversion efficiency.
The number of photoelectrons $N_{pe}$ for a given readout configuration was  determined similarly to that described in \cite{CRADPropEL17}.

All the EL emission components recorded in certain ways can be grouped as follows. There are three basic components for 3PMT+WLS readout: that of the VUV emission of ordinary electroluminescence re-emitted by the WLS; that of the UV emission of NBrS electroluminescence re-emitted by the WLS; that of the visible emission of NBrS electroluminescence recorded directly. There are two components for 1PMT readout: that of the UV and visible emission of NBrS electroluminescence; that of the optical cross-talk from the WLS films in front of the other 3 PMTs. There is one component for SiPM readout: that of the UV, visible and NIR emission of NBrS electroluminescence; the optical cross-talk  from the WLS films is estimated to be negligible. 

For the components where there is a WLS re-emission, we have $PCE=\varepsilon <CE> <QE>$. For the components recorded directly we have $PCE=\varepsilon <QE>$ and $PCE=\varepsilon <PDE>$ for PMT and SiPM readout respectively. Here $\varepsilon$ is the photon collection efficiency; it was calculated using Monte-Carlo simulation procedures. $<CE>$ is the WLS conversion efficiency averaged over the appropriate emission spectrum, namely over that of VUV of ordinary electroluminescence or that of UV of NBrS electroluminescence. 
$<QE>$ and $<PDE>$ are the PMT QE and SiPM PDE averaged over the appropriate emission spectrum and appropriately convolved with the WLS hemispherical, acrylic or UV-acrylic transmittance spectrum given in Fig.~\ref{QEPDE}.  

Regarding the conversion efficiency of the TPB-based WLS, in our previous works \cite{CRADPropEL15,CRADPropEL17} we used $CE=0.58$  around 128 nm \cite{TPB13}, $CE=0.40$ at 300-400 nm \cite{TPB13,TPB13a,TPB96} and $CE=0$ above 400 nm. However, the TPB conversion efficiency at 128 nm has recently been significantly revised towards a decrease \cite{TPB17}: the correction factor reached 2. Therefore we had to measure the WLS conversion efficiency ourselves, in our experimental setup. For this purpose, a cryogenic run was taken to record a scintillation signal in liquid Ar (S1), at zero electric field, induced by vertical tracks of cosmic muons selected by scintillation counters. Knowing, first, the value of the scintillation yield for minimal ionizing particles, of 24.4$\pm$1 eV to produce a VUV photon in liquid Ar \cite{Chepel13,Doke02}, and, second, the photon collection efficiency calculated by us using simulation procedure, we came to the value of $CE=0.49\pm0.02$ which is not far from that used before. Thus in the present work,  we used $CE=0.49$  around 128 nm, $CE=0.40$ at 300-400 nm and $CE=0$ above 400 nm.

Other details of the experimental setup and measurement procedures were described elsewhere \cite{CRADPropEL17}.

\section{Electroluminescence yield: experiment versus theory}

\begin{figure}[!hbt]
	\centering
	\includegraphics[width=0.99\columnwidth,keepaspectratio]{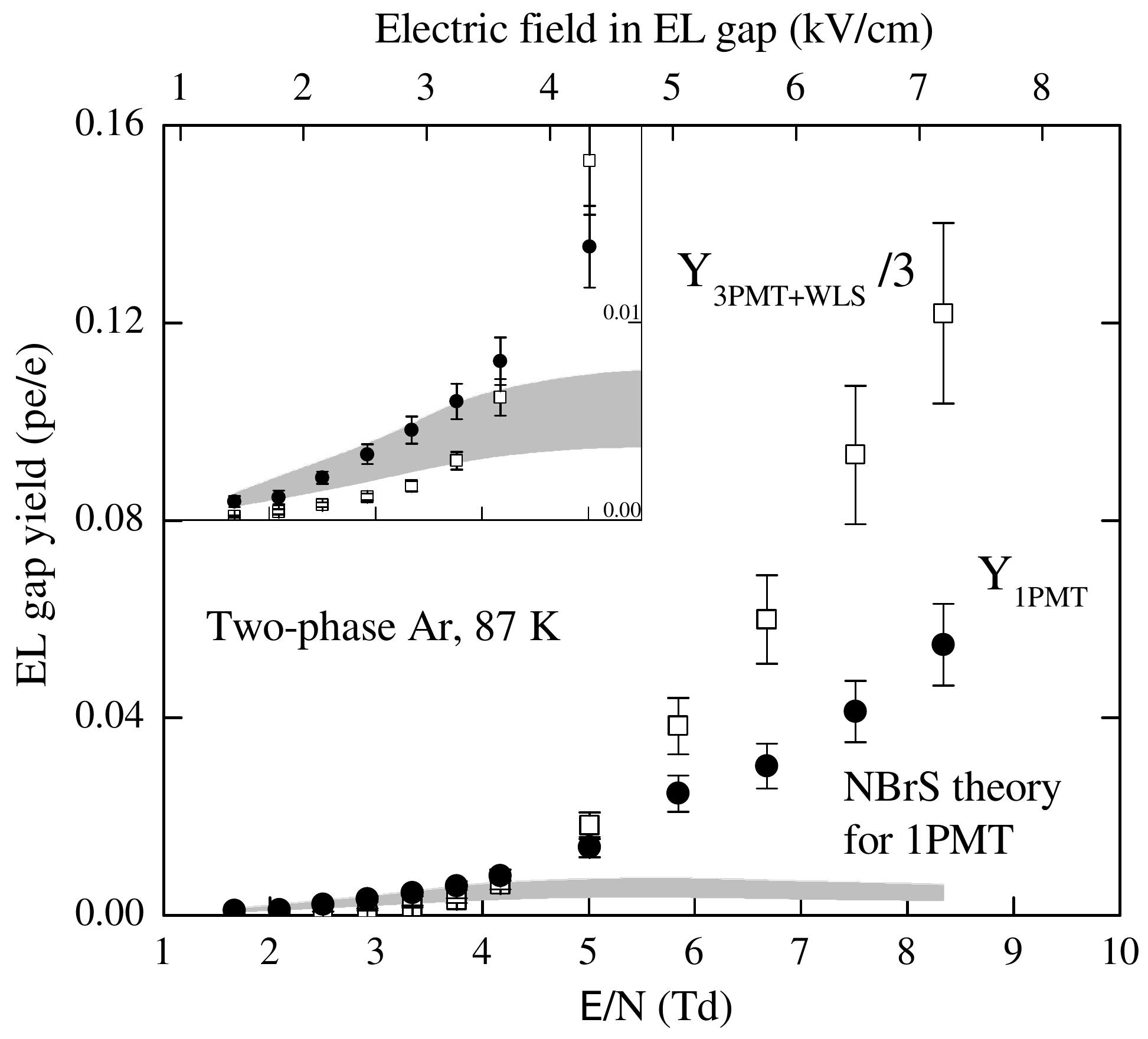}
	\caption{EL gap yield for 3PMT+WLS and 1PMT readout (optical cross-talk due to WLS being subtracted), as a function of the reduced electric field. For convenience of comparison, the 3PMT+WLS yield is divided by 3. The insert shows an enlarged view of the low-field region. The appropriate electric field in gaseous Ar is also shown on the top axis. 
	The prediction of the theory of NBrS electroluminescence for 1PMT readout is also shown. The hatched area indicates the theory uncertainty due to electron energy distribution functions obtained using Boltzmann equation (lower limit) and corrected for mean energy before collisions (upper limit) (see Fig.~\ref{NBrS-yield-field-dep}).}
	\label{EL-gap-yield-PMT-true}
\end{figure}

\begin{figure}[!hbt]
	\centering
	\includegraphics[width=0.99\columnwidth,keepaspectratio]{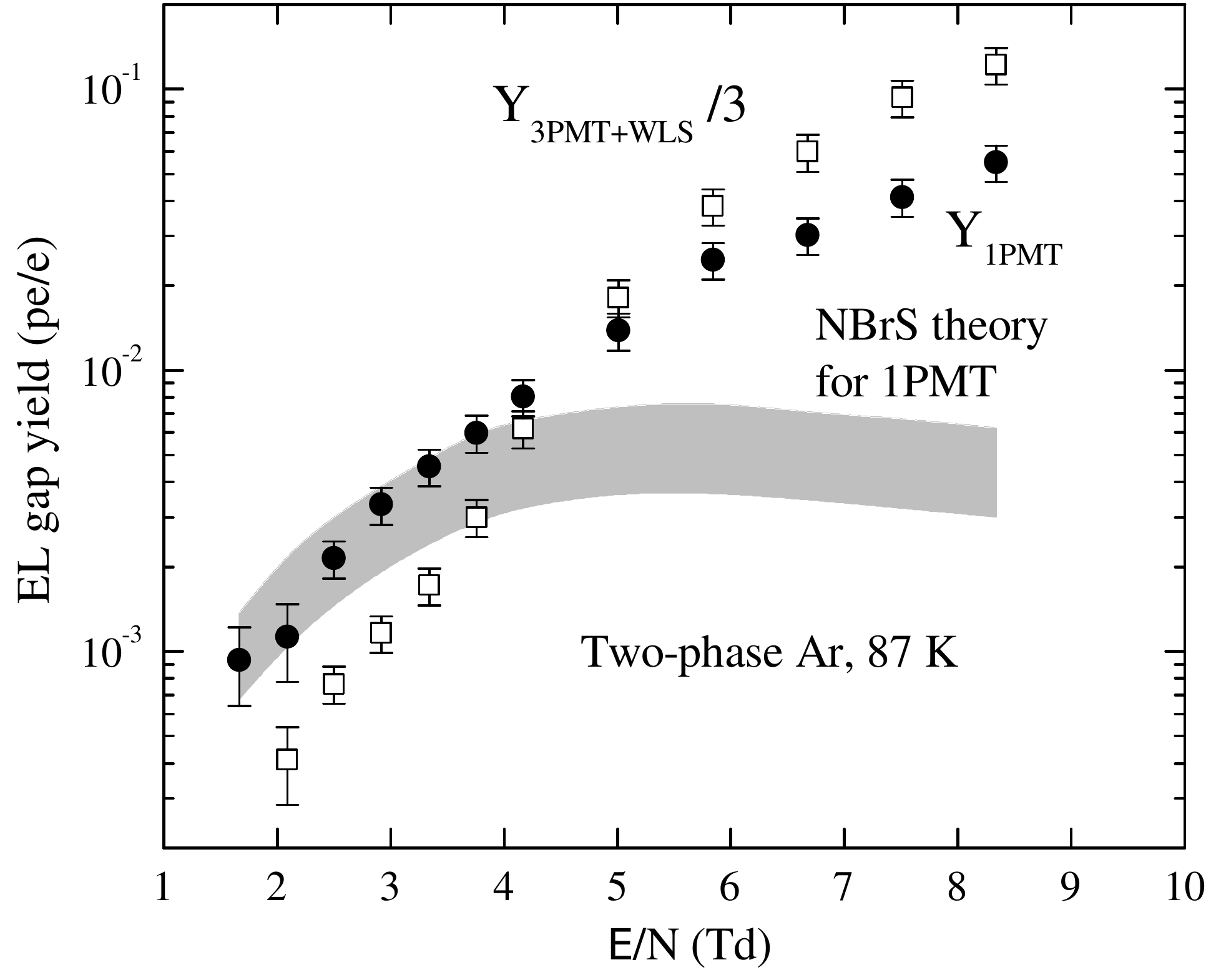}
	\caption{The same as Fig.~\ref{EL-gap-yield-PMT-true}, but in a logarithmic scale.}
	\label{EL-gap-yield-PMT-true-loq}
\end{figure}

Figs.~\ref{EL-gap-yield-PMT-true} and \ref{EL-gap-yield-PMT-true-loq} show the experimental data on the EL gap yield for 3PMT+WLS and 1PMT readout, along with the prediction of the theory of NBrS electroluminescence (for 1PMT readout). Initially the data included the contribution of optical cross-talk to PMT signals due to WLS films of the neighboring PMTs. It was carefully determined by simulation, confirmed in special measurements, and then subtracted to obtain the true EL gap yield. In particular, it was the largest for 1PMT readout, amounting to about 15\% of the initial amplitude at higher electric fields. 

At first glance, Figs.~\ref{EL-gap-yield-PMT-true} confirm the ordinary field dependence of the EL yield: proportional electroluminescence starts at some threshold electric field, of about of 4.0 Td, corresponding to the commencement of atomic excitations, and then linearly increases with the field. However, closer examination reveals two remarkable properties of proportional electroluminescence. First, there is an under-threshold electroluminescence, below the Ar excitation threshold, where the non-VUV component fully dominates. Second, there is an appreciable contribution of the non-VUV component above the threshold. 

These properties are characterized by a relatively strong response of the bare PMT to proportional electroluminescence: its amplitude is as large as 40\% of the amplitude of the PMT with WLS at a higher field, at 8.3 Td. At field values below 4.6 Td (corresponding to that of operating field of the DarkSide-50 experiment \cite{DarkSide15}), the amplitude of the bare PMT starts exceeding that of the PMT with WLS. It should be remarked that the PMT with WLS is 2-3 times less sensitive to NBrS electroluminescence than the bare PMT, since a significant part of the NBrS spectrum (below 400 nm, see Figs.~\ref{NBrS-EL-yield-spectra} and \ref{QEPDE}) is re-emitted by the WLS and thus is suppressed by about a factor of 15-20 \cite{CRADPropEL15}. 

\begin{figure}[!hbt]
	\centering
	\includegraphics[width=0.99\columnwidth,keepaspectratio]{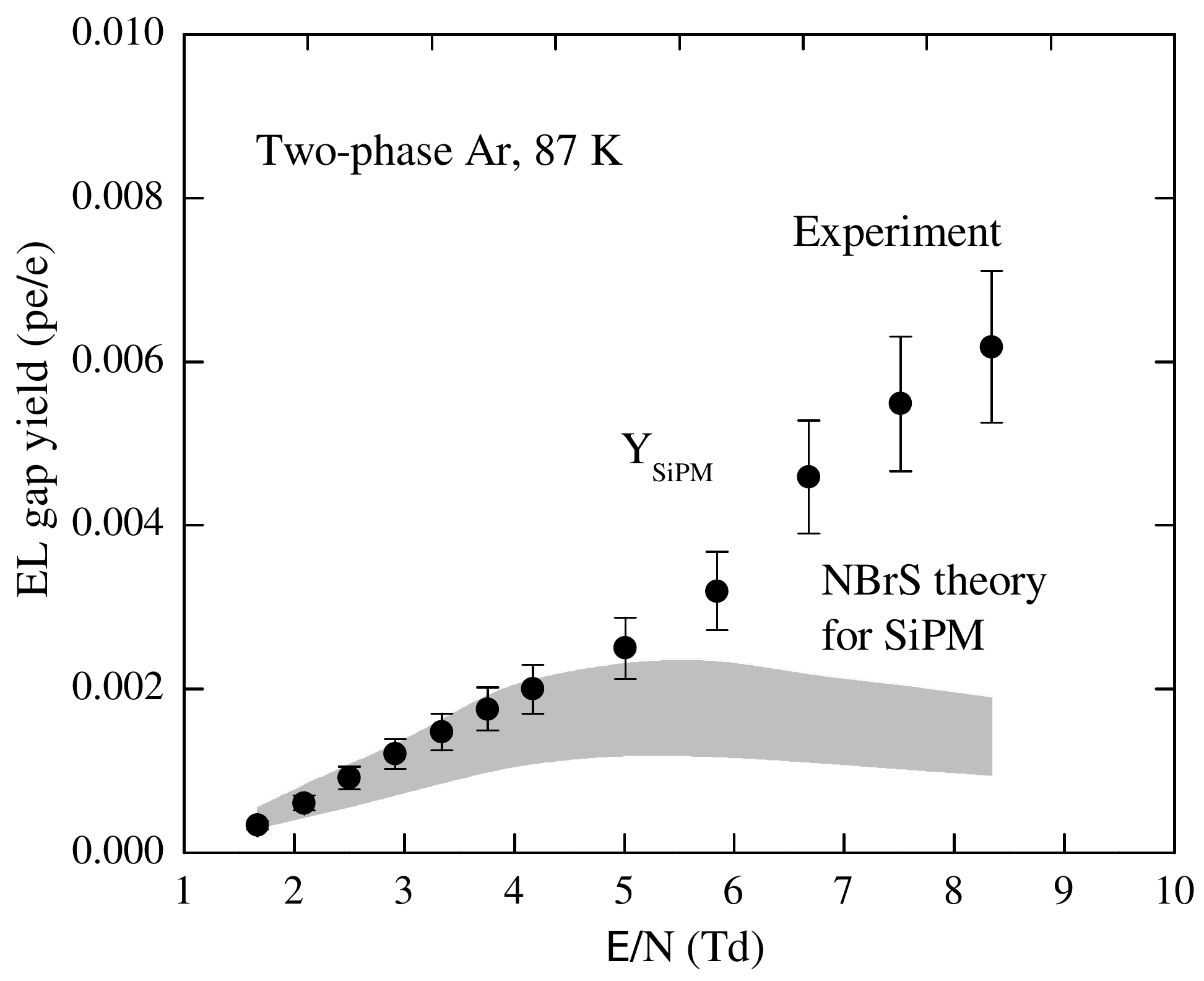}
	\caption{EL gap yield for SiPM readout (at an overvoltage of 3.6 V) measured in experiment and predicted by the theory of NBrS electroluminescence, as a function of the reduced electric field. The hatched area indicates the theory uncertainty due to electron energy distribution functions obtained using Boltzmann equation (lower limit) and corrected for mean energy before collisions (upper limit) (see Fig.~\ref{NBrS-yield-field-dep}). }
	\label{EL-gap-yield-SiPM}
\end{figure}

Fig.~\ref{EL-gap-yield-SiPM} showing the EL gap yield for SiPM readout, fully confirms the results on the non-VUV component obtained using 1PMT readout: electroluminescence below the Ar excitation threshold and the constantly increasing EL yield above the threshold.

One can see from Figs.~\ref{EL-gap-yield-PMT-true},  \ref{EL-gap-yield-PMT-true-loq} and \ref{EL-gap-yield-SiPM} that the non-VUV component is well described by the NBrS theory (within the theoretical uncertainty) at lower electric fields, below the Ar excitation threshold at 4.0 Td. At higher fields, above the threshold, the theory quickly diverges from the experiment. Back in the 80s, it was suggested how to eliminate such a discrepancy \cite{DeMunari71,DeMunari84,Dallacasa80}: one should take into account the specific character of electron scattering in the vicinity of Feshbach resonances. Firstly, one should account for possible electron trapping at Feshbach resonance energies \cite{DeMunari71,DeMunari84}, which leads to the enrichment of the high-energy tail of the electron energy distribution function.  Secondly, as discussed in Section 3, it was theoretically demonstrated that the NBrS yield at resonance might be significantly enhanced \cite{Dyachkov74,Dallacasa80}. Relying on these hypotheses, we adopt here the concept that all the data on non-VUV component are those induced by NBrS electroluminescence.

\begin{figure}[!hbt]
	\centering
	\includegraphics[width=0.99\columnwidth,keepaspectratio]{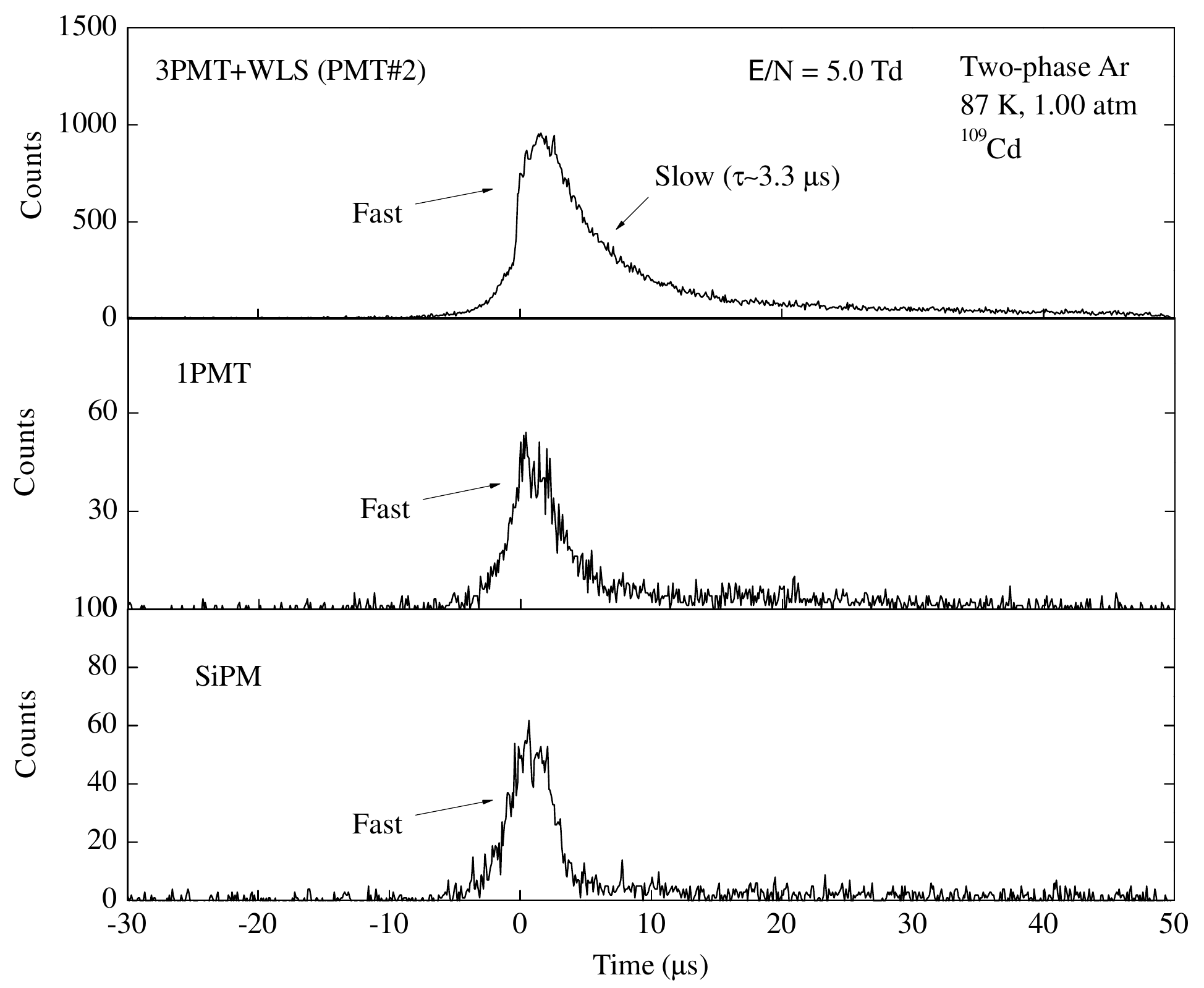}
	\caption{Time spectra of single pe peaks in 3PMT+WLS (PMT\#2), 1PMT and SiPM signals, obtained using peak-finder algorithm, at 5.0 Td, when irradiated with 88 keV gammas from $^{109}$Cd source.  The spectra reflect the pulse-shapes of the appropriate signals. The trigger was that of simple edge, provided by 3PMT+WLS signals. }
	\label{Time-spectra}
\end{figure}

This concept is confirmed by the fact of the photon emission under the Ar excitation threshold in a wide spectral range, which actually is a signature of the NBrS effect. Another signature of the NBrS effect is its fast emission nature, which is revealed in time measurements, above the Ar excitation threshold. In particular it is manifested when comparing the pulse shapes of the signals from the EL gap for all three readout configurations, at a reduced field of 5.0 Td.  Fig.~\ref{Time-spectra} shows time spectra of single photoelectron (pe) peaks for 3PMT+WLS (PMT\#2), 1PMT and SiPM readout. The time spectra reflect the pulse-shapes of the appropriate signals (though they may not completely correspond to the latter because of the possible pe peak merging).  At 5.0 Td, the pulse-shape for 3PMT+WLS readout looks very similar to that of the DarkSide-50 experiment having a close operating field value (4.6 Td) \cite{DS-S2-PulseShape}. One can see that it has a fast and slow component, the latter having a time constant of $\sim$3 $\mu$s and thus being attributed to the VUV emission of the triplet excimer state Ar$^{\ast}_2(^{3}\Sigma^+_u)$ \cite{ArXeN2Proc17}.  The fast component for 3PMT+WLS readout is combined from the non-VUV emission due to NBrS effect and the VUV emission of the singlet excimer state Ar$^{\ast}_2(^{1}\Sigma^+_u)$, according to the current understanding of the EL mechanism. For 1PMT and SiPM readout, the pulse-shape is different: it has basically the fast component only. This is the signature of the NBrS emission in the non-VUV range. The pulse width at half maximum of this fast component is equal to the electron drift time across the EL gap at a given electric field, thus confirming the statement that the signals are produced in the gas in the EL gap (and not in the liquid). The detailed study of time characteristics of the two-phase argon detector will be presented in our next paper.

\begin{figure}[!hbt]
	\centering
	\includegraphics[width=0.99\columnwidth,keepaspectratio]{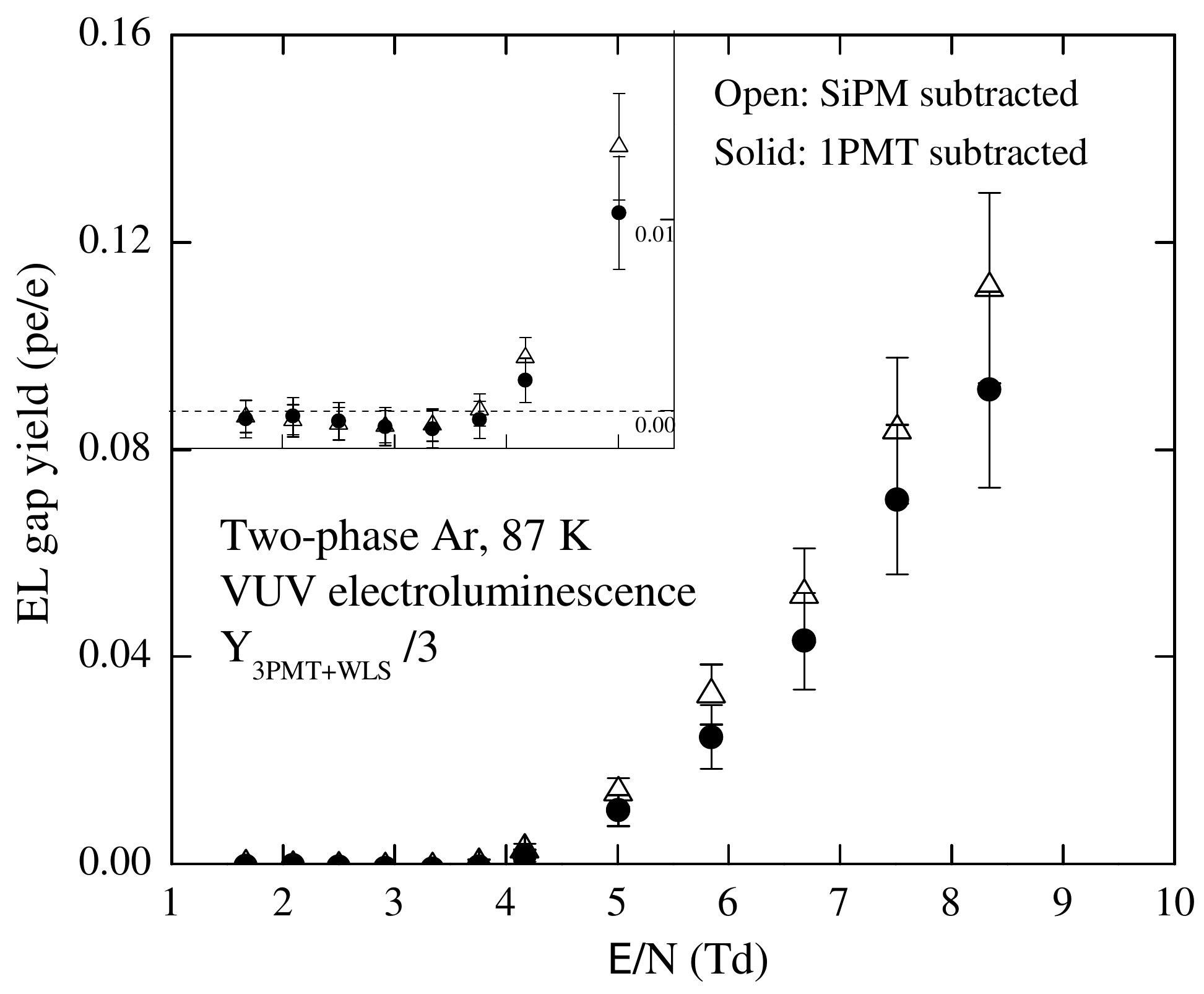}
	\caption{EL gap yield for 3PMT+WLS readout for ordinary (VUV)  electroluminescence  as a function of the reduced electric field, where the non-VUV component is subtracted using the 1PMT and SiPM data, its emission spectra at given electric fields being provided by the NBrS theory. The insert shows an enlarged view of the low-field region.}
	\label{EL-gap-yield-VUV}
\end{figure}

\begin{figure}[!hbt]
	\centering
	\includegraphics[width=0.99\columnwidth,keepaspectratio]{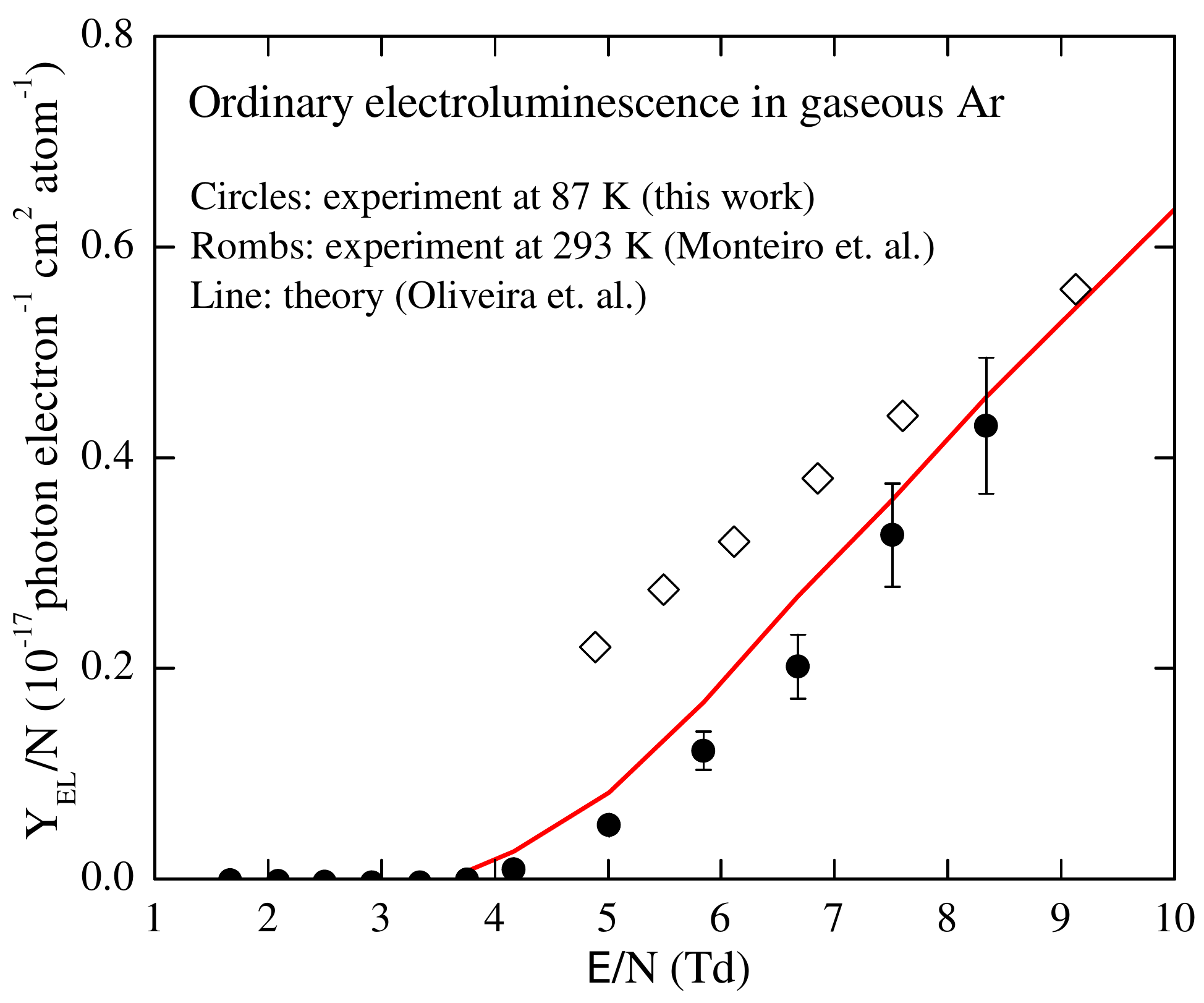}
	\caption{Reduced EL yield for ordinary electroluminescence (in the VUV)  as a function of the reduced electric field, obtained in this work from the experimental data, where the non-VUV component (averaged over 1PMT and SiPM readout) was subtracted using emission spectra of the NBrS electroluminescence model. For comparison, the yields in gaseous Ar in the VUV obtained at 293K experimentally (Monteiro et al. \cite{ArELExp08}) and theoretically (Oliveira et al. \cite{ArELTheory11}) are presented.}
	\label{Ord-EL-yield-exp-theory}
\end{figure}

Adopting the paradigm of the NBrS nature of the non-VUV component, we can determine the true EL gap yield for ordinary (VUV)  electroluminescence from the 3PMT+WLS data, where the non-VUV component is subtracted using the 1PMT and SiPM data, Here the shapes of emission spectra of the non-VUV component were provided by the NBrS theory (Fig.~\ref{NBrS-EL-yield-spectra}). 
The result is shown in Fig.~\ref{EL-gap-yield-VUV}. It can be seen that in the absence of under-threshold radiation, the  threshold of VUV electroluminescence, at 4.0 Td, is much more clearly defined. One can also see that the 1PMT and SiPM data lead to compatible results. Taking the average between them, we arrive at the final Fig.~\ref{Ord-EL-yield-exp-theory}, showing the reduced EL yield for ordinary (VUV) electroluminescence obtained in this work and compared to the yields at room temperature obtained experimentally \cite{ArELExp08} and theoretically \cite{ArELTheory11}. Fig.~\ref{Ord-EL-yield-exp-theory} demonstrates a convincing agreement between the EL yield of ordinary electroluminescence determined here in the two-phase mode, and that of calculated by the theory using microscopic approach \cite{ArELTheory11}.

\section{Discussion and possible applications of NBrS electroluminescence}

First of all, we shall try to prove that there is no rational mechanism other than that of NBrS, which would be able to explain electroluminescence in the non-VUV range and under the Ar excitation threshold. Indeed, physics of electroluminescence is an exact science: all the reaction rate constants of possible EL mechanisms in the presence of impurities are known \cite{ArXeN2Proc17}, allowing to rule out their effect on electroluminescence at the given impurity content limits. In our experiment this limit was below 1 ppm for N$_2$, Xe and other non-electronegative impurities in total, while for electronegative (oxygen-equivalent) impurities it was below 5 ppb (corresponding to the electron lifetime in liquid Ar exceeding 100 $\mu$s). With such vanishingly small impurity contents, there is only one known mechanism that can explain the EL effect under the Ar excitation threshold, namely that of neutral bremsstrahlung. Above the Ar excitation threshold, these content values are several orders of magnitude smaller than those at which the excitation transfer from Ar species to impurities could take place. In particular for N$_2$ the effective excitation transfer in the gas phase starts at $>$10$^3$ ppm, while for Xe it does not occur at all due to extremely low value of the saturated vapor pressure \cite{ArXeN2Proc17}. Finally, regarding possible effect of dissolving of the TPB-based WLS in liquid Ar reported recently \cite{TPB18}, it does not matter for our case since it acts only in the liquid phase. Moreover, it was shown in \cite{TPB18} that the WLS films composed of non-saturated TPB in polymer matrix, i.e. similarly to that used in our experiment (1 part of TPB per 3 parts of polystyrene \cite{CRADPropEL15}), are resistant to dissolving in liquid Ar. Summarizing, it is difficult to imagine the EL mechanism other than that of NBrS that could coherently explain the observed phenomena in total, including electroluminescence under the Ar excitation threshold, its fast emission nature, the non-VUV emission above the threshold and its specific spectral composition recorded in concert by different spectral devices like that of PMT and SiPM. 

Thereby in this work, we got closer to resolving the problem of proportional electroluminescence in two-phase Ar, identified in \cite{ArXeN2Proc17,CRADPropEL17}. Following \cite{ArXeN2Proc17}, it can be stated that the admixture of N$_2$ to Ar, at a content of the order of 50 ppm, cannot be responsible for the non-VUV component in proportional electroluminescence  observed earlier in \cite{CRADPropEL15,CRADPropEL17}. Most probably it is neutral bremsstrahlung responsible for that. The question still remaining unanswered is that of the EL yield enhancement observed in \cite{CRADPropEL15,CRADPropEL17}. What unites those works is that the measurements were performed with freshly prepared WLS films. Accordingly, a WLS instability might be the explanation of the observed enhancement. Indeed, there are indications that WLS films, first showing high conversion efficiency, can then degrade with a half life varying from ten days to several months, depending on the intensity of light irradiation \cite{WLSAgeing}. 

The important result of this work is that the amplitude of the S2 signal from the bare PMT (without WLS) was shown to be comparable with that of the PMT with WLS, in the absence of optical contact between the WLS and the PMT. The latter condition resulted in that the photon flux is considerably  reduced after re-emission by the WLS, by about a factor of 15-20 \cite{CRADPropEL15,CRADPropEL17}, due to re-emission and total reflection losses. 
The S2 amplitude can be further increased, by a factor of 2, if to replace the bare PMTs with the SiPM-matrices, since the latter have higher PDE and wider range of sensitivity to NBrS spectra. This is confirmed by convolving the spectra from Figs.~\ref{NBrS-EL-yield-spectra} and \ref{QEPDE}. This observation paves the way for direct readout of the S2 signal in two-phase dark matter detectors, using PMTs and SiPM-matrices.

The presence of the NBrS component in proportional electroluminescence may result in suggesting to analyze the S2 pulse-shape in a new way, in particular in two-phase Ar dark matter detectors \cite{DS-S2-PulseShape,DS-Simul}. For example, at 4.6 Td (i.e. just at operating field of the DarkSide-50 experiment) one has to take into account the substantial enhancement of the fast component due to NBrS electroluminescence: based on the analysis of Figs.~\ref{NBrS-spectra-int}, \ref{QEPDE} and \ref{EL-gap-yield-PMT-true}, the NBrS contribution to the total S2 signal recorded by PMTs with WLS is estimated to be about 50\%. In principle, such an enhancement of the fast component can affect the algorithm for decomposing the S2 signal into a fast and slow component, which in turn can affect the determination of the quantities using the fast component, such as the diffusion coefficients in liquid Ar \cite{DS-S2-PulseShape} or z-coordinate fiducialization. 

It should be emphasized that NBrS electroluminescence has a universal character:  it should be present in all noble gases, including He, Ne, Kr, Ar and Xe. That is why we assume that NBrS electroluminescence is present in S2 signals of two-phase Xe detectors \cite{Xenon12,Lux13,Panda14,Red17}, including those of Liquid Hole Multiplier \cite{LHM15}. Presumably it has not been yet observed due to the fact that the S2 signal in Xe is recorded directly using PMTs with quartz windows (i.e. unlike Ar, without losses due to re-emission in WLS), and at higher electric fields, to provide efficient extraction of the electrons from liquid Xe. This left almost no chance for NBrS signal to be observed at the background of a strong main signal: this is evident from Fig.~\ref{NBrS-vs-ord-EL-yield} while comparing the EL yields of ordinary and NBrS electroluminescence at higher electric fields. 

It is not yet clear how large the contribution of NBrS electroluminescence is in avalanche scintillations, which are used in combined THGEM/SiPM \cite{CRADRev12,Bondar13} and THGEM/CCD \cite{Mavrokoridis14} multipliers. So far, direct recording of avalanche scintillations by SiPMs was considered to be provided by the NIR emission of higher-level argon excitation states Ar$^{\ast}(3p^54p^1)$ \cite{ArXeN2Proc17,CRADRev12,NIREL11,NIRELSim}. 

We may also suppose that the NBrS effect can be responsible for proportional electroluminescence observed in liquid Ar and Xe using immersed GEM-like structures \cite{CRADRev12,LHM13,Lightfoot09}. At first glance it might not be the case, since the liquid Ar density at 87 K is 244 times larger than that of gaseous Ar \cite{ArXeN2Proc17}, resulting in appropriate reduction of $\mathcal{E}/N$ values (in the first approximation liquid Ar can be considered as a strongly compressed gas, obeying the $\mathcal{E}/N$ scaling law \cite{Boyle15}). However at a closer look, it turns out that the electric fields in the center of GEM or THGEM holes used in liquid Ar, of 60-140 kV/cm \cite{CRADRev12}, correspond to $\mathcal{E}/N=0.3-0.7$~Td which are not that small. For such reduced electric fields, the theory predicts that NBrS electroluminescence already exists: see Figs.~\ref{NBrS-yield-field-dep} and \ref{NBrS-vs-ord-EL-yield}. It also predicts the linear dependence of the EL yield observed in experiment. Thus, one cannot exclude the NBrS origin of proportional electroluminescence in liquid noble gases. 

It is also worth mention the application of NBrS radiation which is already used in practice to develop a detection technique for ultra-high-energy cosmic rays \cite{AlSamarai16}. Here the NBrS radiation in the radio-frequency range  (to be recored with antennas
on the Earth surface) is emitted by primary ionization electrons left after the passage of the showers in the atmosphere. In the   method developed in \cite{AlSamarai16} to calculate the NBrS signal, Eq. \ref{Eq-sigma-el} was used and non-stationary Boltzmann equation was solved to obtain the energy distribution functions of the primary ionization electrons. 

This phenomenon could in principle explain the weak primary scintillations in liquid Ar in the visible and NIR range, observed earlier by a number of groups \cite{NIREL11,Heindl10,Alexander16},
Indeed, if the secondary scintillation in the VUV in gaseous Ar (S2 signal) is associated with the primary scintillation in the VUV in liquid Ar (S1 signal), why does this not happen with the NBrS scintillation? Following \cite{AlSamarai16}, such primary scintillations in liquid Ar in the visible and NIR range might be explained by neutral bremsstrahlung of the primary ionization electrons.

\section{Conclusions}

In this work we have studied an additional mechanism of proportional
electroluminescence (EL) in gaseous Ar in the two-phase mode, namely that of neutral bremsstrahlung (NBrS) of drifting electrons, acting concurrently with that of ordinary electroluminescence. 
It was shown that the NBrS effect can explain two intriguing observations in EL radiation: that of the substantial contribution of the non-VUV spectral component, extending from the UV to NIR, and that of the photon emission at lower electric fields, below the Ar excitation threshold.
Hence, the NBrS effect has partially resolved the problem of proportional electroluminescence in two-phase Ar, identified in \cite{ArXeN2Proc17,CRADPropEL17}. 

The merit of the present work is that it transformed the idea of NBrS electroluminescence from a hypothesis into a quantitative theory. 
The success of the NBrS theory developed here is that it correctly predicts the absolute value of the EL yield below the Ar excitation threshold. The success of both the experiment and the theory is that the data, obtained by three different readout techniques, namely using PMTs with WLS, bare PMT and SiPM, become  consistent if to use the NBrS electroluminescence spectra. This resulted in that the EL yield of ordinary electroluminescence (in the VUV) has for the first time been correctly determined in gaseous Ar at cryogenic temperature, in the two-phase mode. 

The main practical application of the NBrS effect is a better understanding of the S2 signal and justification for its direct (without WLS) optical readout using PMTs and SiPM-matrices, which may help to develop two-phase dark matter and neutrino detectors of ultimate sensitivity.

\section*{Acknowledgments}

This work consisted of two parts, theoretical and experimental (sections 1-3 and 4-6). The first part was supported by Russian Science Foundation (project no. 16-12-10037); the second part was supported by Russian Foundation for Basic Research (project no. 18-02-00117). It was done within the R\&D program of the DarkSide-20k experiment. 

\section*{References}


\end{document}